\documentclass[11pt,a4paper,notitlepage]{article}
\usepackage[a4paper,top=3cm,bottom=3.5cm,left=3.cm,right=3.cm]{geometry}

\usepackage{amssymb}
\usepackage{booktabs}
\usepackage{amsmath}
\usepackage{graphics}
\usepackage{graphicx}
\usepackage{enumitem}
\usepackage{color}
\usepackage{environ}
\usepackage{tikz}
\usepackage{titling}
\usepackage{blindtext}
\usepackage[square,numbers]{natbib}

\NewEnviron{elaboration}{
	\par
	\begin{tikzpicture}
	\node[rectangle,minimum width=0.9\textwidth] (m) {\begin{minipage}{0.85\textwidth}\BODY\end{minipage}};
	\draw[dashed] (m.south west) rectangle (m.north east);
	\end{tikzpicture}
}

\title{\Huge{\textbf{Spectral properties of the trap\\model on sparse networks}}}
\author{Riccardo Giuseppe Margiotta$^{\mathrm{1, *}}$, Reimer K\"uhn$^{\mathrm{1}}$, Peter Sollich$^{\mathrm{1}}$\\[0.2cm]$\mathbf{\mathrm{1}}$ King's College London, Department of Mathematics, Strand,\\London WC2R
		2LS, United Kingdom\\[0.2cm]$^*$\emph{riccardo\_giuseppe.margiotta@kcl.ac.uk}}
\begin{document}
\maketitle

%////////////////////////////////////////////////////////////////////////////////////////%

\begin{abstract}
One of the simplest models for the slow relaxation and aging of glasses is the trap model by Bouchaud and others, which represents a system as a point in configuration-space hopping between local energy minima. The time evolution depends on the transition rates and the network of allowed jumps between the minima. We consider the case of sparse configuration-space connectivity given by a random graph, and study the spectral properties of the resulting master operator. We develop a general approach using the cavity method that gives access to the density of states in large systems, as well as localisation properties of the eigenvectors, which are important for the dynamics.
We illustrate how, for a system with sparse connectivity and finite temperature, the density of states and the average inverse participation ratio have attributes that arise from a non-trivial combination of the corresponding mean field (fully connected) and random walk (infinite temperature) limits. In particular, we find a range of eigenvalues for which the density of states is of mean-field form but localisation properties are not, and speculate that the corresponding eigenvectors may be concentrated on extensively many clusters of network sites.
\end{abstract}

%////////////////////////////////////////////////////////////////////////////////////////%

\section{Introduction}\label{section1 }

Glasses are disordered materials that do not exhibit the structural periodicity of crystals but nonetheless possess the mechanical behaviour of solids. The most common way of making a glass is by quenching, i.e.\ cooling a viscous liquid so rapidly that crystallisation is avoided. The resulting system is called a \emph{supercooled liquid}. The quench brings the molecules of the material into a configuration where the typical time needed to rearrange them is so long that the structure of the liquid appears frozen. The system falls out of equilibrium in the sense that the relaxation time becomes of the order of the observation time window.
The resulting extremely slow evolution is called \emph{glassy dynamics}, and the transition into the regime of very long relaxation times is referred to as the \emph{glass transition}. Technically this phenomenon is not a real phase transition as there are no discontinuous changes in any physical property. Nevertheless, one can associate a critical temperature $T_{\mathrm{G}}$ to a certain liquid, below which the rate of change of e.g.\ volume due to a change in temperature is comparable to that of a solid. The value of $T_{\mathrm{G}}$ also depends on the rate at which the system is cooled: slower cooling allows the material to fall out of equilibrium at lower temperatures (allowing more time for configurational sampling).

 Several theoretical approaches have been proposed to investigate the nature of the glass transition; the general discussion is presented in a recent review by Biroli and Berthier \cite{Berthier2011} (see also references therein). In spite of a sustained research effort dedicated to this problem, a full understanding of glasses has not been achieved yet. One of the most successful theories (based on a microscopic description) is the mode-coupling theory, which predicts a dynamical arrest in supercooled liquids associated with a power law divergence of the `slow' time scale \cite{Gotze1995, Reichman2005}. Another important class of models in the context of glassy systems is that of \emph{spin glasses}, where one generally starts from a Hamiltonian $\mathcal{H}$ with disordered interactions and derives the thermodynamic properties of the spin system and its dynamical behaviour by averaging over the disorder \cite{Mezard1986, Castellani2005a}.
 
A further useful angle of attack on the glass problem focusses on the dynamics in configuration-space. The energy landscape of a glass is typically very rugged, consisting of many local minima (metastable states) separated by energetic barriers, and a global minimum (the crystalline equilibrium state) that is kinetically extremely difficult to reach. One can then think of this energy landscape as a set of basins of attraction that act as `traps' for the dynamics: during its evolution towards equilibrium, the system jumps between local minima at rates that decrease strongly with decreasing temperature. Based on this picture, several studies have been developed, focusing on various aspects of glassy dynamics in configuration-space. These range from investigations of the potential energy landscape, in particular the structure and distribution of minima and energetic barriers between them \cite{Buchner1999, DeSouza2009, Heuer2008}, to simplified models that describe the evolution between traps at a more phenomenological level \cite{Bouchaud1992, Bouchaud1995, Monthus1996}. 

Interestingly, once the description of the configuration-space dynamics has been simplified to motion among traps without internal structure, it is directly related to the research field of stochastic processes on networks. The structure of the energy landscape and the relative positions of neighbouring minima define a network of allowed transitions: the system can only jump between traps that are linked within this network, i.e.\ between minima that are close in the configuration-space. Therefore methodology and results from network theory \cite{Albert2002, Newman2003} can be applied to understand the phenomenology of glasses. In particular, information about the energy landscape can be used to model the time evolution of the system as a Markov process on the network of minima, with rates depending on the relevant energy barriers. Mathematically, the problem thus turns into solving a master equation for the time-dependent probability distribution that describes the position of the system in configuration-space.

One of the simplest and most successful descriptions that belong to this framework is the trap model by Bouchaud and others \cite{Monthus1996}. The transition rates are assumed to depend on the depth of the departing trap $j$ only, not on the arrival trap $i$, and have the Arrhenius-like form
%@@ have changed this to $r_{ij}$ for consistency with later notation @@
\begin{equation}
r_{ij}=\frac{1}{N} e^{-\beta E_j}
\end{equation}
Here $\beta$ is the inverse temperature, $E_j>0$ is the trap depth and $N$ is the size of the network (the number of traps). Every transition effectively involves activation to the top of the energy landscape, where all the energy barriers are located, and then falling into a new state that is chosen randomly among all the minima. The latter assumption implies that this model postulates a mean field (fully connected) network structure. It is easy to show that, for an exponential density of trap depths $\rho_E(E)=e^{-E}$, a glass transition occurs at finite temperature. More specifically, below $T_{\mathrm{G}}=1$ the equilibrium probability distribution across trap depths becomes non-normalisable. The exponential form of the trap density of states can be motivated from several points of view, e.g.\ the mean-field replica theory of spin-glasses \cite{Mezard1986}, the random energy model \cite{Derrida1981}, or phenomenological arguments in the context of supercooled liquids \cite{Odagaki1995}. Also, following an extreme value statistics argument, one might expect that deep minima of potential energy landscapes are described by the Gumbel distribution, whose tail is indeed exponential \cite{Bouchaud1997}

The simple expression for the transition rates and the fully connected network structure allow one to solve the master equation for the model described above analytically. This is simplest in the Fourier-Laplace domain, from where the behaviour in the time domain can then be extracted straightforwardly \cite{Monthus1996}. Trap models have been also studied in Euclidean space, where the system jumps between the nodes of a regular lattice; see for example \cite{Monthus1996, Rinn2001} or the work by Ben Arous and collaborators \cite{BenArous2006, BenArous2007}. Variants include branching phenomena \cite{Muirhead2016} and walks on positive integers \cite{Croydon2017}, though this is less plausible when modelling configuration-space dynamics. The first extension to a trap model on a network was considered relatively recently by Baronchelli \emph{et al}, who used a simple heterogeneous mean field approximation to study the dynamics. This assumes that the probability to find the system on a certain site only depends on the degree (i.e.\ on the number of adjacent nodes) of the site. It therefore has to postulate that the trap depth at any site is fixed fully by its degree \cite{Baronchelli2009, Moretti2011}. Numerical results do indeed show some correlation between trap depth and degree \cite{Doye2002, Massen2007}, though the relation between the two is far from deterministic.

In this work we extend the analysis of the trap model to dynamics on generic (random) networks with sparse inter-trap connectivity. Compared to \cite{Moretti2011} we develop a more flexible approach to the modelling of glassy configuration-space dynamics that allows an arbitrary (deterministic or stochastic) relation between trap depth and node degree. Within this general scenario we then consider the simplest case where trap depths are uncorrelated with degrees. 

For disordered energy landscapes with sparse connectivity a direct analytical solution of the dynamics is not possible in either frequency or time domain; we therefore tackle the problem via the spectral properties of the master operator, which are key in determining the dynamics of the system. Specifically we calculate the density of states (DOS), which gives the spectra of relaxation rates of the system, and the localisation properties of the eigenvectors, measured using the inverse participation ratio (IPR). We develop a general cavity method for this purpose, leading in the infinite system size limit to an integral equation that can be solved numerically via a population dynamics algorithm. Technically, the approach follows analogous applications of the cavity method to the spectral analysis of symmetric random matrices; see e.g.~\cite{Rogers2008, Rogers2009, Kuhn2015, Metz2010} or \cite{Bordenave2010, Khorunzhy2004} for a rigorous discussion, and \cite{Bordenave2011a, Bordenave2011, Bordenave2017} for related work on heavy-tailed random matrices. Based on the DOS and IPR, we are able to obtain insights into the relevant time scales and time regimes of the system. However, we do not have access to some time-dependent objects like correlation functions, which are the main quantities of interest within the literature of trap models. Our analysis will therefore be different from that of previous works \cite{Bouchaud1992, Bouchaud1995, Monthus1996, Bertin2003, BenArous2006, BenArous2007, Baronchelli2009}, and limited to describing the dynamics in terms of the static quantities mentioned above.

This paper is organised as follows. In  section \ref{section2} we define the general set-up of the problem. In section \ref{section3} we discuss by way of background the localisation of the ground state as a function of the temperature, and summarise the known results for the mean field and random walk limits. In section \ref{section4} we address the general case of trap model dynamics on networks with finite connectivity and at finite temperature, and we propose a simple analytical approximation for the DOS. Also, we explain how the parameter $\varepsilon$ that appears in the evaluation of the DOS can be exploited as a detection tool for localisation transitions within the spectrum of the system. We then use these methods to extract dynamical properties of the trap model on random regular graphs, where all nodes have the same degree.
In section \ref{section5} we extend the analysis to other network topologies including scale-free graphs. Section \ref{section6} summarises our conclusions and outlines perspectives for future work.

%////////////////////////////////////////////////////////////////////////////////////////%
\section{Problem set-up}\label{section2}

The general setting of the problem is the following: we consider a continuous-time Markov process defined on a network of $N$ nodes that represent the energy states accessible by the system, i.e.\ the minima of the potential energy landscape or simply the traps. The starting point is then given by the master equation for the probability distribution $\mathbf{p}(t)=(p_1(t),\ldots,p_N(t))$, where $p_i(t)$ is the probability to find the system in trap $i$ at time $t$:
\begin{equation}\label{Master_equation}
\partial_t \mathbf{p}(t)=\mathbf{M}\mathbf{p}(t)
\end{equation}
The \textbf{master operator} $\mathbf{M}$ has the following structure:
\begin{equation}
M_{ij}=c_{ij}r_{ij} \quad\quad M_{ii}=-\sum_{j\neq i} M_{ji} 
\end{equation}
where $c_{ij}=c_{ji}=1$ if nodes $i$ and $j$ are connected and $c_{ij}=0$ otherwise, also $c_{ii}=0$ (there are no self-loops), and $r_{ij}$ is the transition rate from node $j$ to node $i$. Note that $\sum_j M_{ji} = 0$, which ensures that probability is conserved.
We shall now specify the trap-depth distribution, the transition rates and the network topology. We assume
\begin{enumerate}
\item exponentially distributed energies: $E\sim\rho_E(E)= e^{-E}$, $E\ge 0$;
\item random graph structure: the $c_{ij}$ are sampled from a random graph ensemble with finite connectivity. The simplest case for our purposes is one where all those graphs have equal probability for which each node $i$ is connected to exactly $c$ others; the probability distribution of node degrees $k_i=\sum_j c_{ij}$ is then $p_k=\delta_{c,k}$.  Samples that belong to this ensemble are called \emph{random regular graphs} (RRG). In this work we are interested in the case of $c\geq 3$. This condition ensures that the fraction of nodes outside the giant component vanishes in the large system limit \cite{Bollobas2001}, which also implies that the configuration-space is connected, therefore ergodic. We will develop our theory for  general random graph ensembles where the degree distribution is constrained to some $p_k$; in that case $c$ is defined as the average degree $c=\sum_k k\,p(k)$.

\item Bouchaud transition rates: $r_{ij}= e^{-\beta E_j}/c \equiv r_j$. The total escape rate from node $j$ is defined as $\hat{r}_j = \sum_i c_{ij} r_{ij}$ and can be written in terms of the node degree as $\hat{r}_j=k_j r_j$. We will find it useful to define $\tau_j=(c r_j)^{-1}= e^{\beta E_j}$. This gives the expected waiting time to exit from trap $j$, exactly so for regular graphs and up to a factor $c/k_j$ in the general case. From the energy distribution we obtain that the $\tau_j$ have the distribution
\begin{equation}\label{tau_distribution}
\rho_{\tau}(\tau)=T\tau^{-(T+1)} \quad \text{with} \quad \tau\in\left[1,\infty\right)
\end{equation}
 which implies that the average waiting time $\langle\tau\rangle$ diverges for $T<1$, signalling the occurrence of glassy dynamics and aging.
\end{enumerate}
With the above assumptions the master operator $\mathbf{M}$ is a sparse random matrix. It will be important to bear in mind that two sources of randomness come into play: the disorder in the trap depths $\{E_i\}$ and in the inter-trap connectivity $\{c_{ij}\}$. Accordingly we also have two different notions of distance that are relevant for this model: the distance in energy, i.e.\ the energy difference among the minima, and the distance on the graph structure. As we will see in the following sections, these notions of distance play different roles, depending on the case being studied, with regards to their relevance for the degree of localisation of the eigenstates.

%\noindent 
The formal solution of equation (\ref{Master_equation}) is given by 
\begin{equation}\label{Spectral_decomposition}
\mathbf{p}(t)=\sum_{\alpha}e^{\lambda_{\alpha}t}(\mathbf{w}_{\alpha},\mathbf{p}(0))\mathbf{u}_{\alpha}
\end{equation}
%\begin{equation*}
%\lambda_{\alpha\neq 0} < 0 \quad \lambda_0 = 0 \quad \text{with} \quad \alpha=0,\ldots, N-1
%\end{equation*}
where $\lambda_{\alpha}, \mathbf{w}_{\alpha}, \mathbf{u}_{\alpha}$ are respectively the eigenvalues, left eigenvectors and right eigenvectors of $\mathbf{M}$, indexed by $\alpha=0,1,\ldots,N-1$, and $(\mathbf{w}_{\alpha},\mathbf{p}(0))$ denotes the scalar product between the left eigenvector $\mathbf{w}_{\alpha}$ and the initial probability distribution $\mathbf{p}(0)$. In the following we will refer to $r_\alpha = -\lambda_\alpha$ as the {\em relaxation rates} of the system, and write $r$ for a generic relaxation rate. If the network is connected, there is a single vanishing eigenvalue $\lambda_0=0$. All other $\lambda_\alpha$ must then have negative real part so that the corresponding modes $\mathbf{u}_{\alpha}$ make a contribution to $\mathbf{p}(t)$ that is exponentially suppressed over time. In the long-time limit only $\mathbf{u}_0$ survives, which is the equilibrium Boltzmann distribution of the system associated with the ground state $\lambda_0$. The corresponding left eigenvector is $\mathbf{w}_0 = (1,\ldots ,1)$. So
\begin{equation}
\lim_{t\to\infty}\mathbf{p}(t)=\mathbf{u}_0=\mathbf{p}_{\mathrm{eq}}=\frac{1}{Z}(e^{\beta E_1},\ldots,e^{\beta E_N})
\end{equation}
Within the present formulation the energies are positive as they represent the depth of each trap, so $e^{\beta E_i}$ is the correct Boltzmann weight for node $i$.

The evolution of the probability at finite $t$ depends on the spectral properties of the master operator. In particular, slowly decaying modes govern the long-time behaviour of the system, and solving the master equation amounts to diagonalising $\mathbf{M}$. This operation can be performed analytically only for a few special cases presented in section \ref{section3}. However, information about the spectrum and the localisation properties of $\mathbf{M}$ can still be obtained in the large system limit; we use the \textbf{cavity method} \cite{Mezard1986a} for this purpose. 
This method links $\mathbf{M}$ to the inverse covariance matrix of a Gaussian distribution, and therefore requires the symmetrised form of the master operator
\begin{equation}\label{Master_operator}
\mathbf{M}^{\mathrm{s}}=\mathbf{P}_{\mathrm{eq}}^{-1/2}\mathbf{M}\mathbf{P}_{\mathrm{eq}}^{1/2}
\end{equation}
or in components $M^{\mathrm{s}}_{ij}=r_i^{1/2}M_{ij}r_j^{-1/2}$,
where we have introduced a diagonal matrix $\mathbf{P}_{\mathrm{eq}}$ with $(\mathbf{P}_{\mathrm{eq}})_{ii}=p_i^{\mathrm{eq}}\propto r_i^{-1}$. This transformation preserves the eigenvalue spectrum of $\mathbf{M}$, implying that the associated eigenvalues are real, as $\mathbf{M}^{\mathrm{s}}$ is real and symmetric. We note that the diagonal elements of $\mathbf{M}$ remain unchanged: $(\mathbf{M}^{\mathrm{s}})_{ii}=(\mathbf{M})_{ii}$. The eigenvectors $\mathbf{v}_{\alpha}$ of $\mathbf{M}^{\mathrm{s}}$ are given by $\mathbf{v}_{\alpha} = \mathbf{P}_{\mathrm{eq}}^{-1/2} \mathbf{u}_{\alpha} = \mathbf{P}_{\mathrm{eq}}^{1/2} \mathbf{w}_{\alpha}$. Physically, the symmetry of $\mathbf{M}^{\mathrm{s}}$ means that the dynamics we are considering obeys detailed balance with respect to the Boltzmann steady state. % @@ can't use transpose in last equation, unless you also write the transpose on $\mathbf{v}_{\alpha}$ @@

Our study aims to predict the statistics of the eigenvalues and eigenvectors of $\mathbf{M}^{\mathrm{s}}$. The first quantity of interest is the density of states (DOS), i.e.\ the fraction of eigenvalues lying between $\lambda$ and $\lambda + \mathrm{d}\lambda$, defined as $\rho(\lambda)\mathrm{d}\lambda$ with
\begin{equation}\label{DOS}
\rho(\lambda)= \frac{1}{N}\sum_{\alpha=0}^{N-1} \delta(\lambda-\lambda_{\alpha})
\end{equation}
We average the DOS over random samples, finitely sized, and assume self-averaging in the thermodynamic limit. The DOS is crucial as it defines the time scales $\{|\lambda_{\alpha}^{-1}|\}$ of the dynamics, or, more precisely, it gives the full spectra of relaxation rates $\{r_{\alpha}\}$.  It is essential to keep this in mind as, for consistency, we will present the results in terms of the DOS throughout the paper, and occasionally remind the reader of the simple relation $r_{\alpha} = -\lambda_{\alpha}$.

The second key quantity that we are interested in is the degree of localisation of the eigenstates, which carries information about their ability to contribute to the transport properties of the system across the network: localised modes can only contribute to local probability-flows. The rationale behind this is clear: assuming $p_i(0)=\delta_{ij}$, then if the vectors $\{\mathbf{u}_{\alpha}, \mathbf{w}_{\alpha}\}$ are mostly localised, only a few terms in the sum on the r.h.s.\ of equation (\ref{Spectral_decomposition}) give a significant contribution to the probability distribution $\mathbf{p}(t)$, which should therefore spread only slowly over time away from the initial node $j$. In general, we expect that the ability of the system to explore the configuration-space depends on the degree of localisation of the eigenvectors of $\mathbf{M}$. To quantify this we use the inverse participation ratio (IPR) defined as
\begin{equation}\label{IPR_q}
I_q(\mathbf{v})=\frac{\sum_{i=1}^{N} v_i^{2q}}{(\sum_{i=1}^{N} v_i^2 )^q}\sim N^{-t_q}
\end{equation}
where $\mathbf{v}=(v_1,\ldots,v_N)$ is an eigenstate; if \textbf{v} is normalised the denominator equals unity. The exponent $t_q$
defines the scaling of $I_q$ with $N$. We refer to \cite{Fyodorov2010} for a general introduction to the IPR and related quantities. In what follows we concentrate on the standard IPR with $q=2$. We can distinguish two extreme situations: if the ``mass'' of the eigenstate $\mathbf{v}$ is evenly spread over all the states of the system, namely each element $v_i$ is of order $1/\sqrt{N}$, then the eigenstate is $\textbf{delocalised}$ and $I_2(\textbf{v})=O(1/N)$, $t_2=1$. If instead only a few elements of $\mathbf{v}$ differ from zero, the eigenstate is $\textbf{localised}$ and $I_2(\textbf{v})=O(1)$, $t_2=0$.

Interestingly, the localisation properties of eigenstates defined on random regular graphs (and random matrices) are studied also in the context of quantum many body systems -- with similar terminology and methodology -- where they are linked to the problem of ergodicity and equilibration dynamics \cite{Altshuler1997, Altshuler2016, Facoetti2016}. 

A number of studies have looked instead at the localisation of the {\em time-dependent probability distribution} of trap models on lattices. Particularly interesting is the 1D case, which exhibits \emph{dynamical localisation} where localisation properties differ between the aging regime and the final Boltzmann distribution \cite{Bertin2003}. Flegel and Sokolov analysed this phenomenon using the spectral properties of the master operator \cite{Flegel2014}; they trace the non-equilibrium value of the IPR during aging back to the eigen\emph{vector} statistics, while the eigen\emph{value} statistics only make a minor contribution. Dynamical localisation is also discussed in the context of statistical mechanics of trajectories \cite{Ueda2017}, which represents another interesting approach to describing the glass transition in terms of configuration-space evolution.

\begin{figure}[htbp]
\centering
\includegraphics[height=5.5cm]{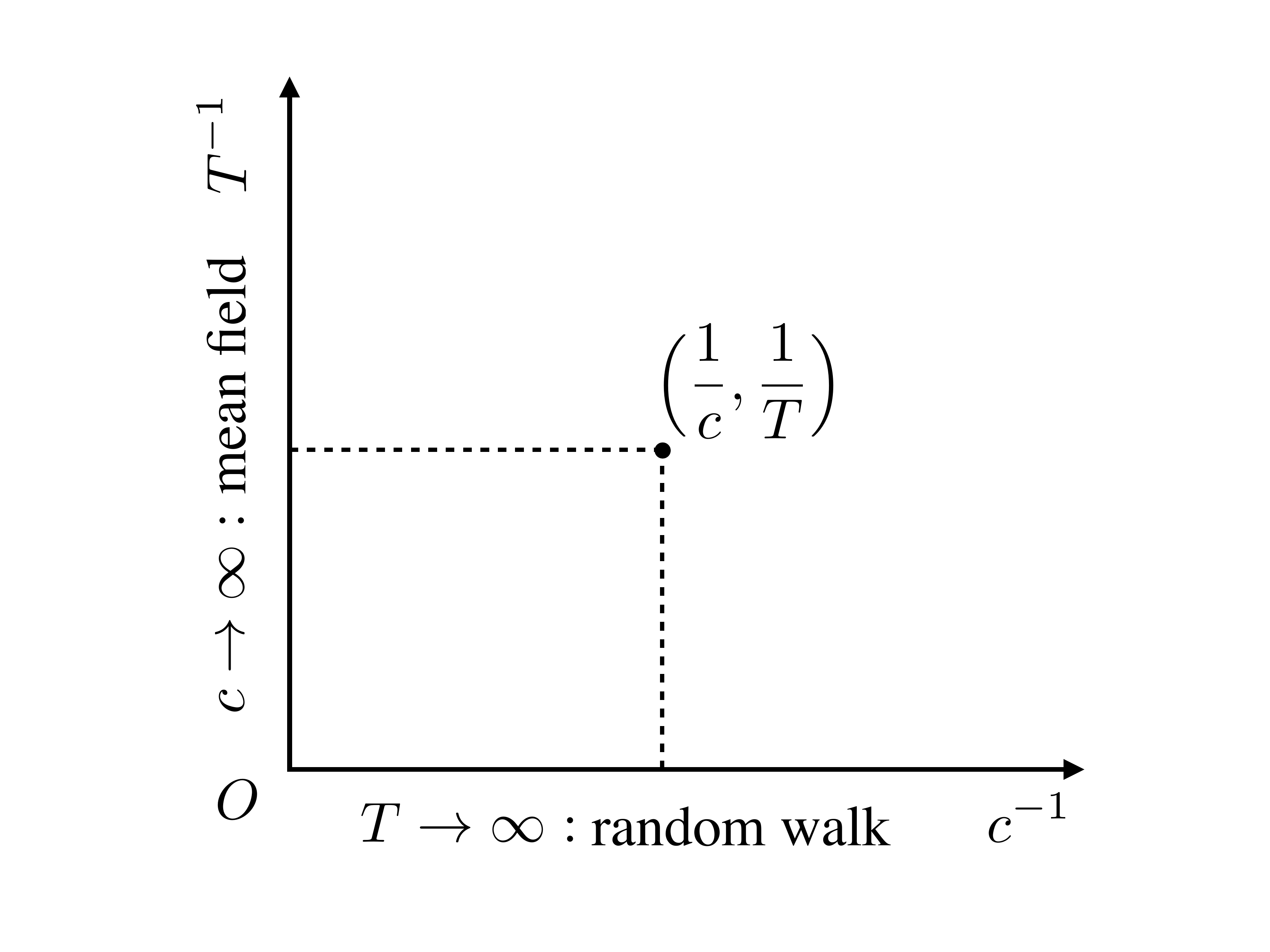}
\caption{Infinite temperature ($x$-axis) and infinite mean connectivity ($y$-axis) limits for the Bouchaud trap model on a network. The point at $(1/c,1/T)$ represents the general case of finite connectivity and finite temperature. Note that we only consider graphs with a giant connected component, which imposes a minimal value of $c$ (e.g. $c=1$ for Erd\"os-R\'enyi graphs \cite{Albert2002}) so that the horizontal axis has a finite range.}\label{fig:Tvsc.pdf}
\end{figure}

The model we study, which is a Markov process on a random graph with Bouchaud transition rates, is described by two main parameters: the temperature $T$ and the mean connectivity $c$. As depicted in figure \ref{fig:Tvsc.pdf}, there are two obvious limits that can be considered: the $\textbf{mean field}$ (MF) $c\to\infty$ limit, where the network structure becomes trivial and only the disorder in energy is present, i.e.\ there are ``glassiness effects'' only, and the $T\to \infty$ limit, where the trap depths become irrelevant and the system effectively performs a $\textbf{random walk}$ (RW) among neighbouring traps. The point shown in the $(1/c, 1/T)$ plane in figure \ref{fig:Tvsc.pdf} represents our model with finite connectivity, at finite temperature. This general case can be thought of as a combination of the two limiting situations of mean field and random walk. In this work we illustrate how, for a system with finite $c$ and $T$, quantities such as the DOS and the average IPR of eigenstates have attributes that arise by a non-trivial combination of the corresponding MF and RW limits.

\section{Ground state and limiting cases}\label{section3}

\subsection{Ground state: $\lambda=0$}

\begin{figure}[htbp]
\centering
\includegraphics{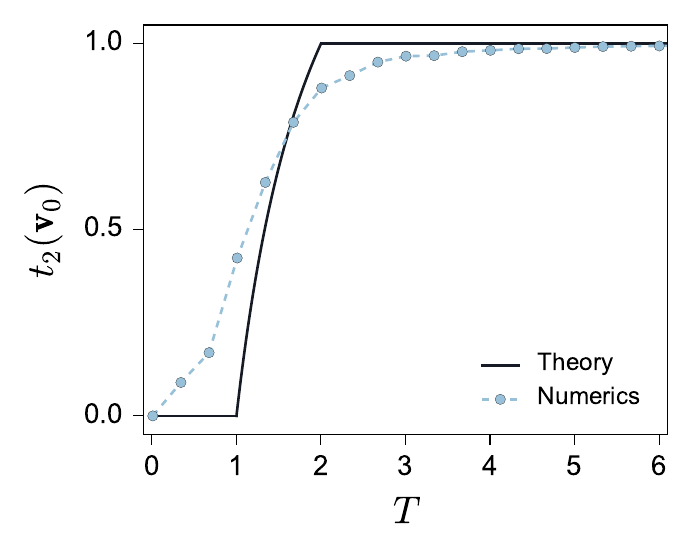}
\caption{Ground state exponent $t_2(\mathbf{v}_0)$ (black) predicted by (\ref{Ground_state_IPR}), and numerical average $\bar{t}_2 (\mathbf{v}_0)$ (dashed line) taken across $M=100$ ground state realisations of size $N=1000$.}\label{fig:GstateAv}
\end{figure}

The eigenvector $\mathbf{u}_0$ represents the equilibrium probability distribution of the system,
$\mathbf{p}_{\mathrm{eq}} = \lim_{t\to \infty} \mathbf{p}(t)$. This is independent of the network structure and its statistics depend only on the energy distribution $\rho_E(E)$. One can assess the degree of localisation of the equilibrium distribution via the IPR of either $\mathbf{u}_0$ or its symmetrised analogue $\mathbf{v}_0$. 
Explicitly, these are proportional to 
\begin{equation}\label{Equilibrium_distribution}
\mathbf{u}_0=\mathbf{p}_{\mathrm{eq}}\propto (e^{\beta E_1}, \ldots, e^{\beta E_N}) 
\quad\quad 
\mathbf{v}_0=\mathbf{p}_{\mathrm{eq}}^{\mathrm{s}} \propto (e^{\beta {E_1}/{2}}, \ldots, e^{\beta {E_N}/{2}})
\end{equation}
The localisation of the ground state depends on whether the Boltzmann weights are concentrated on the deepest traps or not. Since the energies are randomly allocated to the vertices of the network, its topology will not affect the IPR of the equilibrium distribution; the distance in energy is the only relevant one here. From the definition of the IPR, we get for the ground state of the symmetrised master operator
\begin{equation}\label{Ground_state_IPR}
I_2(\mathbf{v}_0) = \frac{\sum_i e^{2\beta E_i}}{(\sum_i e^{\beta E_i})^2} \sim \frac{N\int^{N^{\beta}}_1 d\tau \tau^{1-T}}{(N\int^{N^{\beta}}_1 d\tau \tau^{-T})^2}\\[0.5cm] \simeq \begin{cases} N^{-1} \quad \text{if}\quad T>2  \\ N^{-2 + 2/T} \quad \text{if}\quad 1<T<2  \\ N^0 \quad \text{if}\quad T<1 \end{cases}
\end{equation}
where the cutoff $N^{\beta}$ derives from the extreme value statistics of the distribution $\rho_{\tau}$ \cite{Bouchaud1990}: the $k$ largest waiting times of $N$ samples $\tau_{N+1-k}<\ldots<\tau_{N}$ fall in the range $[\tau_{N+1-k},\infty)$, therefore the fraction $k/N$ is of the order of the area under $\rho_{\tau}$ over this range, which for $k=1$ gives $1/N \simeq \tau_N^{-T} = \tau_{\text{max}}^{-T}$.
From (\ref{Equilibrium_distribution}), the result for the non-symmetrised version is the same except for the replacement of $T$ by $T/2$.
We focus on the symmetrised case as this is the most sensible from the random matrix perspective that we use. The symmetrised eigenvectors are also the natural objects to appear in our cavity approach, which starts from a (complex) Gaussian distribution and hence requires a symmetric covariance matrix as input (see section \ref{section4} and references therein). While the symmetrised eigenvectors do not describe the Markov process that obeys equation (\ref{Master_equation}), away from the ground state the symmetrisation is not expected to affect their localisation properties. In other words, a localised/delocalised symmetrised eigenvector should stay localised/delocalised also in its (either left or right) unsymmetrised form. We refer to appendix \ref{appendixE} for further discussion and data showing that qualitative localization statistics for $\mathbf{u}_{\alpha}$, $\mathbf{w}_{\alpha}$ and $\mathbf{v}_{\alpha}$ are the same except in the small finite-size region of the crossover towards the ground state. We also note that the IPR of the symmetric ground state coincides with the measure of ground state localisation considered in previous works \cite{Derrida1994, Bouchaud1997, Ueda2017}. Finally, using symmetrised eigenvectors to calculate the IPR has an additional benefit: the characteristic temperature where the IPR ceases to be of $\mathcal{O}(1)$ coincides with the glass transition temperature that is known from the dynamics. Indeed, according to equation (\ref{Ground_state_IPR}), the ground state $\mathbf{v}_0$ is localised below the glass transition $T_{\mathrm{G}}=1$, delocalised for $T>2$, and has an intermediate behaviour for $1<T<2$. Figure \ref{fig:GstateAv} shows the exponent $t_2(\mathbf{v}_0)$ of this prediction compared with the average $\bar{t}_2 (\mathbf{v}_0) = \langle -\ln I_2(\mathbf{v}_0)/\ln N\rangle$
of data taken from direct diagonalisations of the symmetrised master operator (hereafter labelled ``numerics" in the plots). Note that in the limit $N\to\infty$  the IPR is of order unity for $T<1$, and drops to zero above, so that the intermediate temperature region ($1<T<2$) should also be regarded as delocalised. In the localised regime, an infinite-$N$ calculation
shows that the $O(1)$ value of the IPR is given explicitly by $I_2(\mathbf{v}_0)=1-T$ for $T<1$ \cite{Bouchaud1997, Derrida1994}, dropping to zero at $T=1$ in agreement with our result. 

\subsection{Random walk limit: $T\to \infty$}

In the infinite temperature limit the dynamics is only affected by the graph topology, i.e. there are network effects only and the differences in energy depth among traps become immaterial. In this case the master operator coincides with its symmetrised form, and it simplifies to
\begin{equation}
M_{ij}=\frac{c_{ij}}{c}-\delta_{ij}
\end{equation}
where the first term on the right hand side is the off-diagonal contribution (because $c_{ii}=0$).
Given that $\mathbf{M}=c^{-1}\mathbf{A}-\mathbf{I}$ is directly related to the adjacency matrix $\mathbf{A}$, its DOS can be deduced where that of $\mathbf{A}$ is known.
For the case of a random regular graph, one obtains the DOS for $N\to\infty$ as a shifted and scaled Kesten-McKay law \cite{McKay1981}:
\begin{equation}\label{RW_DOS}
\rho(\lambda)=\frac{c}{2\pi}\sqrt{4\,\frac{c-1}{c^2}-(\lambda+1)^2}\Big/\Big(1-(\lambda+1)^2\Big)
\end{equation}
which can alternatively be derived using e.g.\ the cavity construction explained below. 
For this graph ensemble all the eigenvectors are delocalised with high probability \cite{Dumitriu2012}. These results are illustrated in figure \ref{fig:MF_RW}-left. 

The dynamics in this case has no glassy features, all local waiting times equal unity so that jumps occur at a constant rate, and the average distance from the initial node grows linearly with time. This is true because, at every jump, the particle has $c-1$ outward paths, and only one inward path pointing towards the starting node, thus the motion effectively resembles a 1D biased random walk. 

\subsection{Mean field limit: $c\to \infty$}

\begin{figure}[t]
\centering
\includegraphics{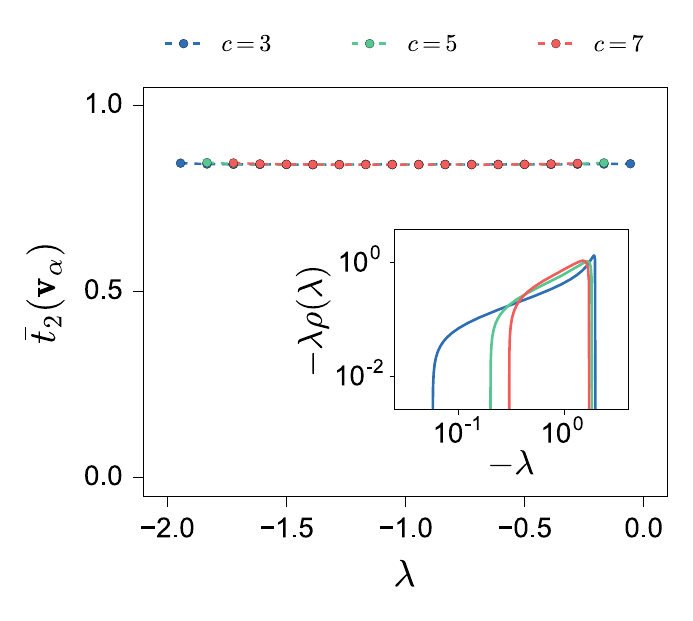}\includegraphics{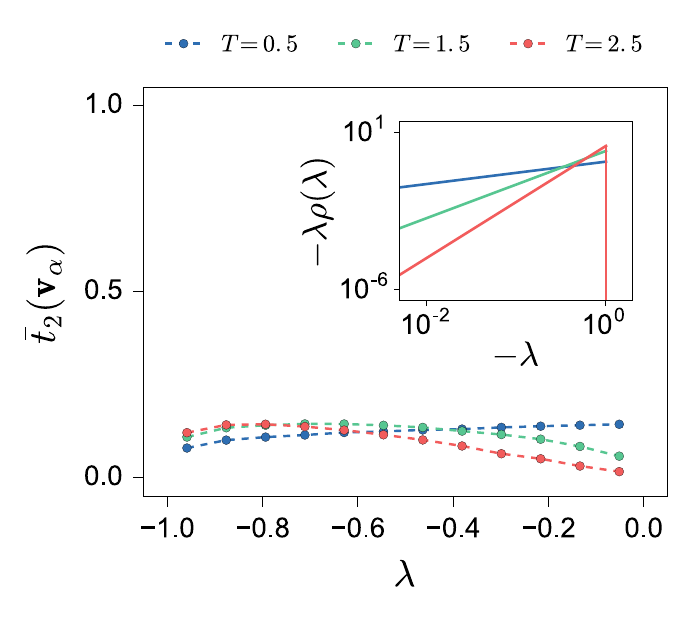}
\caption{Left: Infinite temperature limit for random regular graphs with mean connectivity $c=3, 5, 7$.
Main plot: Average $\bar{t}_2 (\mathbf{v}_{\alpha})$
from direct diagonalisation, with averaging performed both across $M=100$ random graphs and within $\lambda$-bins centred on the symbols. For the finite system size $N=1000$ used, $\bar{t}_2$ is close to but has not yet reached its asymptotic value $1$. Inset: DOS
for $N\to\infty$, given by equation (\ref{RW_DOS}) and plotted as density of $\ln(-\lambda)$ to show the full range; the factor $-\lambda$ appearing on the label of the $y$-axis is the Jacobian of the transformation $\lambda \to \ln(-\lambda)$. We recall that the quantity $-\lambda$ represents the relaxation rate of the system, so this plot can equivalently be read as the density of $\ln (r)$, plotted against $r$ on a logarithmic $x$-axis. Right: Analogous plot for the infinite connectivity limit $c\to\infty$, where $\bar{t}_2\approx 0$ indicates localised eigenvectors, and the DOS is a power law given by (\ref{DOS_MF}).}\label{fig:MF_RW}
\end{figure}

In the infinite $c$ limit the master operator reduces to that of the mean field (fully connected) case. There is no notion of space and only the distance in energy is relevant for the degree of localisation of the eigenstates. For such a fully connected system of size $N$  we have (approximating $c=N-1\approx N$, which is immaterial for $N\to\infty$)
\begin{equation}\label{MF_master_operator}
M_{ij}=\frac{e^{-\beta E_j}}{N}(1-\delta_{ij})-e^{-\beta E_j}(1-\frac{1}{N})\delta_{ij}
\end{equation}
The eigenvalue equation in this case reads
\begin{equation}
\frac{1}{N}\sum_{j\neq i} e^{-\beta E_j}u_{\alpha, j} - e^{-\beta E_i}(1-\frac{1}{N})u_{\alpha, i}=\lambda_{\alpha} u_{\alpha,i}
\end{equation}
This can be written as 
\begin{equation}\label{MF_eigenvalue_equation}
\frac{1}{N}\sum_{j} e^{-\beta E_j}u_{\alpha, j} - e^{-\beta E_i}u_{\alpha, i} = \lambda_{\alpha} u_{\alpha,i}
\end{equation}
and, as the first term is independent of $i$, one has
\begin{equation}\label{MF_eigenvector_element}
u_{\alpha, i} \propto (\lambda_{\alpha} + e^{-\beta E_i})^{-1}
\end{equation}
Similarly, for the symmetrised case one obtains
\begin{equation}\label{MF_eigenvector_element_symmetric}
v_{\alpha, i} \propto \frac{e^{-\beta E_i/2}}{\lambda_{\alpha} + e^{-\beta E_i}}
\end{equation}
Note that for $\lambda_{\alpha}=0$ these expressions recover the equilibrium distribution (\ref{Equilibrium_distribution}) as they should. The solution (\ref{MF_eigenvector_element}) is in agreement with \cite{Bovier2005} where the spectral properties of the mean field trap model are discussed extensively. From (\ref{MF_eigenvector_element}, \ref{MF_eigenvector_element_symmetric}) we expect the IPR to be of order one for all $\lambda \neq 0$, for either their symmetrised or non-symmetrised forms. A simple argument for this localisation result in mean field is presented in appendix \ref{appendixA}; we note here only that the eigenvector entries decay as a power law with energy difference to the ``centre'' of the eigenvector at  $e^{-\beta E_i} \approx-\lambda_{\alpha}$ (the ``centre'' has to be understood here as defined on the energy axis). Returning to the eigenvalues, the condition for $\lambda_\alpha$ follows from equations (\ref{MF_eigenvalue_equation}) and (\ref{MF_eigenvector_element}) as
\begin{equation}
\sum_j\frac{e^{-\beta E_j}}{N(\lambda_{\alpha}+e^{-\beta E_j})} =1
\end{equation}
which implies that there is an eigenvalue in each interval $(-\tau_{i}^{-1}, -\tau_{i+1}^{-1})$, assuming that the energies are ordered so that $E_i<E_{i+1}$ (recall that $\tau_i=e^{\beta E_i}$). Therefore in the large $N$ limit the DOS is given by
\begin{equation}
\rho(\lambda)=\int \mathrm{d}\tau\, \rho_{\tau}(\tau)\, \delta(\lambda+\frac{1}{\tau})
\end{equation}
which gives
\begin{equation}\label{DOS_MF}
\rho(\lambda)=T(-\lambda)^{T-1}
\end{equation}
for $-1<\lambda<0$. These results are shown in figure \ref{fig:MF_RW}-right. Note that the eigenvalue condition (19) and the interleaving of eigenvalues between the (negative) inverse trap lifetimes can also be seen from the fact that the mean field master operator (\ref{MF_master_operator}) is a diagonal matrix with a rank one perturbation \cite{Ran2012}, as all elements in each column are the same except for those appearing on the diagonal.

Interestingly, all modes remain localised at any finite temperature, while glassiness manifests itself only for $T<1$, and even then only for the ground state. This circumstance has to be attributed to the slow decay of the mass of MF eigenvectors away from their localization centre, which does not impair the mobility of the particle. This agrees with the intuition that, in the absence of spatial structure, the particle is always able to reach any node of the network in finite time, as long as the average trapping time is finite, i.e.\ for $T>1$. 

\section{Finite connectivity and finite temperature}\label{section4}

\subsection{The cavity method}

We now turn to the main contribution of our work: moving on from the two limiting scenarios discussed above, we study the general case of finite connectivity and finite temperature where the distance on the graph structure and separation between trap energies are both relevant. We do this by means of the cavity method, exploiting the fact that the random graphs we consider become locally treelike in the large $N$ limit. The master operator in this case has the general structure (\ref{Master_operator}). \emph{In what follows we omit the index} ``$\mathrm{s}$" and consider the symmetrised master operator only. The DOS of the matrix $\mathbf{M}$ can be written in terms of the resolvent $\mathbf{G}(\lambda_{\varepsilon})$ as
\begin{equation}\label{DOS_resolvent}
\rho(\lambda)=\lim_{\varepsilon \to 0} \frac{1}{\pi N}\sum_{i=1}^{N} \text{Im}\,G_{ii}(\lambda_{\varepsilon})
\end{equation}
where
\begin{equation}\label{Resolvent}
\mathbf{G}(\lambda_{\varepsilon})=\big[\lambda_{\varepsilon}\mathbf{I} - \mathbf{M}  \big]^{-1}
\end{equation}
Here
$\lambda_{\varepsilon} = \lambda - \mathrm{i}\varepsilon$, with $\varepsilon$ a small, positive quantity, while $\mathrm{i}$ is the imaginary unit; $\mathbf{I}$ indicates the $N\times N$ identity matrix. To derive equation (\ref{DOS_resolvent}) one replaces the delta distributions in (\ref{DOS}) with Lorentzians of width $\varepsilon$ and takes the limit $\varepsilon\to 0$; this explains the origin of the small imaginary term in $\lambda_{\varepsilon}$. For a detailed description we refer to the original work of Edward and Jones \cite{Edwards1976}.
We define the complex Gaussian measure $P(\mathbf{x})$ as 
\begin{equation}
P(\mathbf{x})\propto e^{-\frac{\mathrm{i}}{2}\mathbf{x}^T \mathbf{G}^{-1}\mathbf{x}}= e^{-\frac{\mathrm{i}}{2}\sum_{i,j}(\lambda_{\varepsilon}\delta_{ij}-M_{ij}) x_i x_j}
\end{equation}
with $\mathbf{x}=(x_1,\ldots ,x_N)$. The diagonal entries of the resolvent are then given by the local variances
\begin{equation}\label{Resolvent_entries}
G_{jj}= \mathrm{i}\int \mathrm{d}x_j\, x_j^2 P(x_j)
\end{equation}
where $P(x_j)$ is the marginal distribution
\begin{equation}
P(x_j)=\int \prod_{k\neq j} \mathrm{d}x_k\, P(\mathbf{x})
\end{equation}

To make further progress we recall that the off-diagonal terms of the symmetrised master operator are $M_{jk}=c_{jk}r_j^{1/2}r_k^{1/2}$, while the diagonal terms are $M_{jj}=-\sum_k c_{kj}r_j$. This gives  
\begin{equation}\label{Complex_gaussian_distribution_x}
P(\mathbf{x})\propto e^{-\frac{\mathrm{i}}{2}[\lambda_{\varepsilon}\sum_j x_j^2 - \sum_{jk} c_{jk} (-r_j x_j^2 +  r_j^{1/2} r_k^{1/2} x_j x_k )]}
\end{equation}
Symmetrising $r_j x_j^2$ to $(r_j x_j^2 + r_k x_k^2)/2$
allows the term in brackets in the last sum to be written as a complete square. 
Equation (\ref{Complex_gaussian_distribution_x}) can be further simplified by the change of variables $y_j=x_jr_j^{1/2}$, which has the benefit of confining the disorder from the transition rates $r_j$ to the local terms:
\begin{equation}
P(\mathbf{y})\propto e^{-\frac{\mathrm{i}}{2}[\lambda_{\varepsilon}\sum_j y_j^2/r_j + \frac{1}{2}\sum_{jk} c_{jk} (y_j-y_k)^2]}
= \prod_j e^{-\frac{\mathrm{i}}{2}\lambda_{\varepsilon}\sum_j y_j^2/r_j}
\prod_{(jk)\in \mathcal{G}}
e^{-\frac{\mathrm{i}}{2}
(y_j-y_k)^2}
\end{equation}
where the last product runs over all distinct edges of the graph $\mathcal{G}$ defined by the inter-trap connectivity $\{c_{ij}\}$.

The core of the cavity approach is to decompose $P(\mathbf{y}) $ into the factors involving a given node $j$, and the remaining factors. The latter define the cavity 
graph $\mathcal{G}^{(j)}$, where node $j$ and all its connections have been removed from $\mathcal{G}$, and a corresponding cavity distribution denoted $P^{(j)}(\cdot)$. This leads to the following equation for the marginal distribution $P(y_j)$:
\begin{equation}\label{Magrinal_distribution_y}
P(y_j)=e^{-\frac{\mathrm{i}}{2}\lambda_{\varepsilon}y_j^2/r_j}  \int \mathrm{d}\mathbf{y}_{\partial j}\, e^{-\frac{\mathrm{i}}{2}\sum_{k\in \partial j}(y_j-y_k)^2}\,P^{(j)}(\mathbf{y}_{\partial j})
\end{equation}
where $P^{(j)}(\mathbf{y}_{\partial j})$ denotes the (complex) probability distribution of the variables $\{y_k\}$ on the nodes that are neighbours of $j$ on the graph $\mathcal{G}$.
The cavity method is based on the assumption that the joint distribution $P^{(j)}(\mathbf{y}_{\partial j})$ factorises on $\mathcal{G}^{(j)}$ as
\begin{equation}\label{Cavity_assumption}
P^{(j)}(\mathbf{y}_{\partial j})=\prod_{k\in \partial j}P^{(j)}(y_k)
\end{equation}
This is exact if the original graph $\mathcal{G}$ is a tree, because the cavity graph $\mathcal{G}^{(j)}$ then consists of disconnected branches. Sparse graphs do contain loops, but these have an average length of order $\ln(N)$ \cite{Albert2002}. Intuitively, as the total number of nodes in the $k$-th coordination shell is $c^k$, these will typically be distinct as long as $c^k\ll N$. Conversely, different sub-trees rooted in the neighbourhood of $j$ can be connected to a common site, hence producing a loop, if $c^k = \mathcal{O} (N)$, or $k = \mathcal{O} (\ln (N))$. In the large $N$ limit these graphs therefore become locally treelike and the factorisation (\ref{Cavity_assumption}) will again become exact: conditional on a given node, the branches rooted at that node become independent of each other. Equation (\ref{Magrinal_distribution_y}) then simplifies to
 \begin{equation}\label{Magrinal_distribution_y_simple}
 P(y_j)=e^{-\frac{\mathrm{i}}{2}\lambda_{\varepsilon}\frac{y_j^2}{r_j}}  \prod_{k\in\partial j}\int \mathrm{d}y_k \, e^{-\frac{\mathrm{i}}{2}(y_j-y_k)^2}P^{(j)}(y_k)
 \end{equation}
Similarly one can show for the marginals of the cavity distribution around node $j$
\begin{equation}\label{Magrinal_cavity_distribution_y}
P^{(j)}(y_k)=e^{-\frac{\mathrm{i}}{2}\lambda_{\varepsilon}\frac{y_k^2}{r_k}} \prod_{l\in \partial k\setminus j}\int  \mathrm{d}y_l \, e^{-\frac{\mathrm{i}}{2}(y_l-y_k)^2}P^{(k)}(y_l)
\end{equation}
where $\partial k\setminus j$ indicates the neighbourhood of node $k$ excluding node $j$. As all distributions involved are zero mean Gaussians, also the marginals must be of this form, i.e.\ 
\begin{equation}\label{Marginal_ansatz_distribution_y}
P^{(j)}(y_k)=\sqrt{\frac{\omega_k^{(j)}}{2\pi}}e^{-\frac{1}{2}\omega_k^{(j)}y_k^2}
\quad\quad
P(y_j)=\sqrt{\frac{\omega_j}{2\pi}}e^{-\frac{1}{2}\omega_j y_j^2}
\end{equation}
We follow statistical terminology and call the $\omega$, which are inverse variances, precisions \cite{Bishop2006}. Their real part must be positive in order to preserve normalisability of the corresponding Gaussians. 

In terms of the precisions and using $r_j^{-1} = \tau_j c$,  equations (\ref{Magrinal_distribution_y_simple}), (\ref{Magrinal_cavity_distribution_y}) become 
\begin{equation}\label{Precisions}
\omega^{(j)}_k=\mathrm{i}\lambda_{\varepsilon} \tau_k c + \sum_{l\in \partial k \setminus j} \frac{\mathrm{i}\omega_l^{(k)}}{\mathrm{i}+\omega_l^{(k)}}
\quad \quad
\omega_j=\mathrm{i}\lambda_{\varepsilon} \tau_j c + \sum_{k\in \partial j} \frac{\mathrm{i}\omega_k^{(j)}}{\mathrm{i}+\omega_k^{(j)}}
\end{equation}
as derived in more detail in appendix \ref{appendixB}. 
The equations for the cavity precisions form a closed set $\{\omega_k^{(j)}\}$ that can be solved iteratively.
Note that for an actual tree, no iteration is required as the equations can be solved recursively by working inwards from the leaves. Once the cavity precisions are known, the marginal precisions $\{\omega_j\}$ can be deduced.
Finally, from (\ref{Resolvent_entries}) and (\ref{Precisions}) one obtains the diagonal entries of the resolvent: 
\begin{equation}\label{Resolvent_entries_precision}
G_{jj}=\frac{\mathrm{i} \tau_j c}{\omega_j}
\end{equation}
The factor $\tau_j c = r_j^{-1}$ arises here from the transformation from $y_j$ back to $x_j$.

Given a specific realisation of the disorder, i.e.\ for a single instance of the matrix $\mathbf{M}$, we know the rates $\{r_i\}$ and the connections $\{c_{ij}\}$, therefore ({\ref{Precisions}}) can be solved iteratively starting from a suitable initial condition. The eigenvalue spectrum of the system is finally given by (\ref{Resolvent_entries_precision}) and (\ref{DOS_resolvent}). We will refer to this procedure as the \emph{single instance cavity method}.

In the large $N$ limit (\ref{Precisions}-left) turns into a self-consistent equation for the \emph{distribution} $p (\omega)$
of the cavity precisions -- here to keep the notation clean we drop the superscript indicating the cavity graph. For the simplest case of a random regular graph this reads 
\begin{equation}\label{Zeta}
p(\omega)=\int \mathrm{d}\tau\,\rho_{\tau}(\tau)\prod_{l=1}^{c-1}\mathrm{d}\omega_{l}\,p(\omega_{l})\,\delta(\omega - \Omega_{c-1})
\end{equation}
where 
\begin{equation}\label{Omega_c}
\Omega_{a} = \Omega_{a}(\{\omega_l\},\tau)=\mathrm{i}\lambda_{\varepsilon} \tau c + \sum_{l=1}^{a}\frac{\mathrm{i}\omega_{l}}{\mathrm{i}+\omega_{l}}
\end{equation}
The intuition here is that because $p(\omega)$ is a distribution resulting from the solution of the equations (\ref{Precisions}) for the cavity precisions, updating the precision on a randomly chosen edge of the graph with the r.h.s.\ of (\ref{Omega_c}) does not change the distribution.
Technically, one  assumes here that the distribution of $G_{jj}$, and consequently $p(\omega)$, is  self-averaging in the limit $N\to\infty$. In the case of a general random graph, the only change is an additional average over the number of neighbours $k$ of a randomly chosen edge, with the appropriate probability weight $kp_k/c$; see appendix \ref{appendixB} for details.

 A numerical solution for $p(\omega)$ at any given $\lambda_{\varepsilon}$ can be obtained using a \emph{population dynamics} algorithm \cite{Mezard2001}. The basic idea is to represent the distribution $p(\omega)$ with a population of $N_{\text{p}}$ cavity precisions $\mathcal{P}=(\omega_1,\ldots,\omega_{N_{\text{p}}})$. One starts with a certain initial condition and lets the population evolve according to the update rule given by the delta function in (\ref{Zeta}). Once equilibrated, the histogram of $\mathcal{P}$ should give an approximation of $p(\omega)$. In summary the algorithm works as explained in the following box:
\begin{center}
\begin{elaboration}
{$\,$ \\[+0.1cm]}
\centering{Population Dynamics Algorithm}\\[+0.2cm]

\begin{enumerate}
\item Start with an initial (complex) population $\mathcal{P}=(\omega_1,\ldots,\omega_{N_{\text{p}}})$.
\item Pick $c-1$ random elements $\{\omega_l\}$ from $\mathcal{P}$ and a sample $\tau$ from $\rho_{\tau}$.
\item Replace a random element of the population with $\Omega_{c-1}(\{\omega_l\},\tau)$.
\item Repeat 2 and 3 until equilibration is reached.
\end{enumerate}
{$\,$ \\[-0.8cm]}
\end{elaboration}
\end{center}
Finally, we use (\ref{DOS_resolvent}) and (\ref{Resolvent_entries_precision}) to write the DOS as an average over the distributions $p(\omega)$ and $\rho_{\tau}$
\begin{equation}\label{DOS_average_precisions}
\rho(\lambda) = \lim_{\varepsilon \to 0} \lim_{N\to\infty}\frac{1}{\pi N}\sum_{i=1}^{N} \text{Im}\,G_{ii}(\lambda_{\varepsilon})=\lim_{\varepsilon \to 0} \frac{1}{\pi} \text{Re}\Big\langle \frac{\tau c}{\Omega_c(\{\omega_l\},\tau)} \Big\rangle_{\{\omega_l\},\tau}
\end{equation}
where the $\{\omega_l\}$ are sampled from the population of cavity precisions converged to equilibrium.

We next discuss the relative merits of single instance cavity method versus population dynamics, and the influence of $\varepsilon$.
The single instance method allows us to find the spectrum of (large) sparse symmetric matrices $\mathbf{M}$, under the cavity approximation of factorisation in each cavity graph. In terms of computational cost this method is in principle much faster than direct diagonalisation because one only has to find the $O(N)$ cavity precisions, typically from a number of iterations of the cavity equations that does not grow with $N$.  However, one still has to store all the information on the disorder $\{\tau_i, c_{ij}\}$ and, as is true generally with the cavity technique, one obtains little information about the eigenstates. The calculation also has to be repeated across a suitably fine grid of $\lambda$-values in order to find the spectrum.

In choosing the $\lambda$-grid, one has to bear in mind that the general approach replaces the $N$ delta-functions in (\ref{DOS}) by Lorentzians of width $\varepsilon$, therefore $\varepsilon$ \emph{is the ``resolution" that we have on the lambda axis}. To catch all eigenvalues of a single instance, one therefore requires a grid spacing in $\lambda$ of order $\varepsilon$ or smaller. Conversely, for a fixed $\lambda$-grid, $\varepsilon$ has to be chosen larger than the grid spacing, otherwise the chance of hitting all eigenvalues becomes too low to obtain accurate results.

In practice, we always perform the cavity iterations themselves with $\varepsilon=\varepsilon_0\to 0$ (specifically we set $\varepsilon_0\sim10^{-300}$) so that the resulting cavity precisions are not affected by the width of the Lorentzians. The required nonzero $\varepsilon$ ($\gg\varepsilon_0$) is then applied only in the evaluation of the average (\ref{DOS_average_precisions}), i.e.\ in the measurement step. 
This makes it easy to explore the effect of changes in $\varepsilon$, without having to solve the cavity equations afresh.
From (\ref{Marginal_ansatz_distribution_y}, \ref{Precisions}) one sees that using $\varepsilon\neq \varepsilon_0$ to calculate the marginal precisions is equivalent to adding $\varepsilon-\varepsilon_0$ to the inverse variance of each $x_j$. This provides a regularization for the case where the variance calculated using $\varepsilon_0$ is close to imaginary because the chosen $\lambda$ has hit an eigenvalue.
 
In contrast to the single instance approach, the population dynamics algorithm is designed to give the DOS of infinitely large systems. There is no need to keep track of the disorder because of self-averaging, and we only have to let the population equilibrate. The eigenvalue spectrum becomes densely populated, typically showing a continuous part referred to as the \emph{bulk}. This means that we are always able to compute the DOS over this region, even with $\varepsilon_0\sim 10^{-300}$. A common feature of (sparse) random matrices is that the states covering the bulk are in fact delocalised (or \emph{extended}), and localisation (\emph{Lifshitz}) tails are present at the edges of the spectrum \cite{Khorunzhiy2006, Bapst2011, Metz2010, Biroli1999}. The values of $\lambda$ where these \emph{localisation transitions} occur are called \emph{mobility edges}. Pure points, i.e.\ isolated eigenvalues \cite{Reed1980}, do sometimes occur within the bulk of the spectrum, as is the case for e.g.\ sparse adjacency matrices with varying node degrees \cite{Kuhn2008}. In the following sections we refer to the density of all the states of the system as the \emph{total} DOS (tDOS), obtained by the population dynamics algorithm with $\varepsilon$ small but finite, and to the density of the extended states only as the \emph{extended} DOS (eDOS), obtained with $\varepsilon$ effectively equal to zero. 

\subsection{Total DOS via population dynamics}

\begin{figure}[htbp]
\centering
\includegraphics{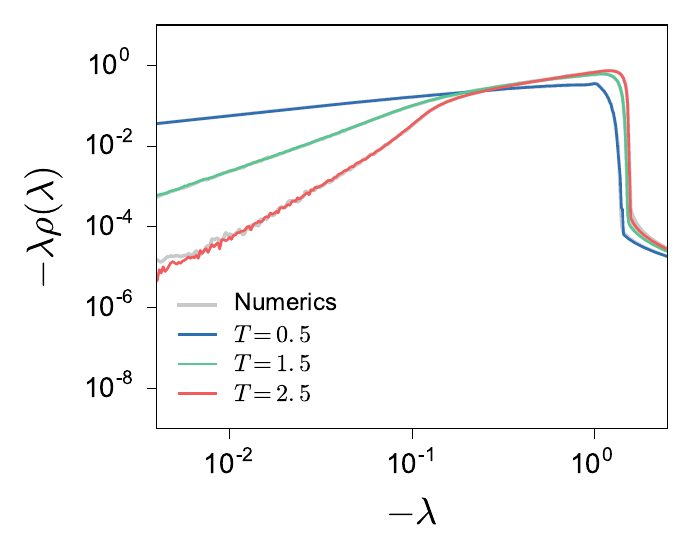}\includegraphics{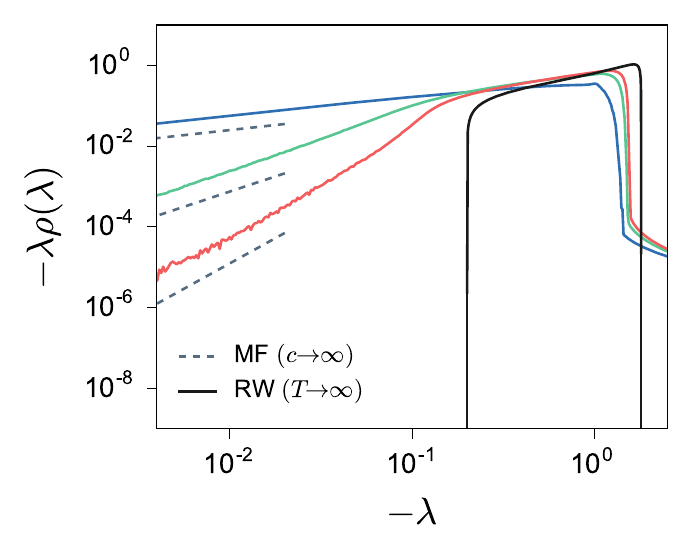}
\caption{Total DOS for mean connectivity $c=5$ and temperatures $T=0.5,1.5,2.5$. Left: predictions from cavity method (population dynamics, evaluated using $\varepsilon\sim10^{-4}$ and population size $N_{\mathrm{p}} = 2500$) compared to direct diagonalisation spectra (grey) for networks of size $N=1000$, with statistics taken across $M=10^4$ graph samples. Right: cavity predictions for total DOS compared with MF and RW limits, respectively given by (\ref{RW_DOS}) and (\ref{DOS_MF}).}\label{fig:DOS_lims}
\end{figure}

We next present the results for the total DOS of a trap model on a random regular graph with connectivity $c=5$. Figure \ref{fig:DOS_lims}-left shows the results obtained using the population dynamics algorithm compared with data from direct diagonalisation of the master operator for finite $N$ (labelled ``numerics"). The agreement across the entire $\lambda$ range is clearly very good. In figure \ref{fig:DOS_lims}-right we include the MF and RW-limits of the DOS for comparison; recall that the quantity $r=-\lambda$ represents the relaxation rate of the system, so the plots showing $-\lambda\rho(\lambda)$ vs $-\lambda$ can equivalently be read as the density of $\ln (r)$, plotted against $r$ on a logarithmic $x$-axis. We observe that the small $|\lambda|$ tails (the slow modes governing the long-time dynamics) follow the MF trend (blue dashed lines), showing the same power law exponent asymptotically. Conversely, fast modes (large $|\lambda|$) show a non-linear DOS which originates primarily from the Kesten-McKay law (RW limit). We note here that because of the MF tail, one expects systems of finite size to have a spectral gap that scales with $N$ as in mean field \cite{Bovier2005}, so the second largest eigenvalue should be bounded from above by $-\tau_{\text{max}}^{-1}\sim-N^{-\beta}$; a detailed analysis of the $N$-scaling of the spectral gap, however, is beyond the scope of the present work. Note that at the highest temperature $T=2.5$, the small $|\lambda|$ tail of the DOS shows larger statistical uncertainties because of finite size effects: in direct diagonalisation, finite-sized matrices only rarely have eigenvalues in this region; similarly population dynamics sampling runs of finite length produce only a limited number of samples contributing to the slow mode regime. 

To understand the structure of the DOS in more qualitative terms, we can perform a simple (high $T$) analytical approximation: we take one cavity iteration at finite temperature starting from the infinite temperature solution. This means that only the local disorder is taken into account when computing the DOS, i.e.\ the central node receives its messages from $c$ neighbours belonging to an infinite temperature cavity network. A similar idea, called the \emph{single defect approximation}, has been used to explain localisation phenomena arising from topological disorder in random lattices \cite{Biroli1999, Semerjian2002}. In the large $T$ and $N$ limits, where all nodes become equivalent, the cavity precision distribution $p(\omega)$ becomes delta-peaked on the value $\bar{\omega}$ that  solves (\ref{Zeta}), i.e.
\begin{equation}
\bar{\omega}=\Omega_{c-1}(\{\bar{\omega}\},1)
\end{equation}
The approximated total DOS is then evaluated as in (\ref{DOS_average_precisions})
\begin{equation}\label{DOS_approximation}
\rho^{\mathrm{A}}(\lambda)=\lim_{\varepsilon\to 0}\frac{1}{\pi}\text{Re}\Big\langle \frac{\tau c}{\Omega_c(\{\bar{\omega}\},\tau)} \Big\rangle_{\tau}
\end{equation}
The average can be performed analytically, as detailed in appendix \ref{appendixC}. Figure \ref{fig:Approx}-left shows the resulting first order approximation $\rho^{\mathrm{A}}(\lambda)$ against the DOS obtained by direct diagonalisation. One observes that the approximation is in remarkably good agreement with the numerical data in the region of slow (MF-like) modes, though even in the RW-like regime it is qualitatively correct. We can iterate the scheme to obtain higher order approximations: to the second order we perform two cavity iterations at finite temperature starting from the infinite temperature solution, and so on. The second order approximation is then given by
\begin{equation}
\rho^{\mathrm{2A}}(\lambda)=\lim_{\varepsilon\to 0}\frac{1}{\pi}\text{Re}\Big\langle \frac{\tau c}{\Omega_c(\{\Omega_{c-1}(\{\bar{\omega}\},\tau_l)\},\tau)} \Big\rangle_{\{\tau_l\},\tau}
\end{equation}
This average cannot be carried out analytically but is straightforward to  perform by sampling from the distribution of waiting times. Figure \ref{fig:Approx}-right shows the first and second order approximations on a linear scale. One gets a slightly better result with the second order approximation $\rho^{\mathrm{2A}}$ in the large $|\lambda|$-region, though not yet a quantitative  match to the full cavity predictions. In the small $|\lambda|$ tail we find (not shown here) that there is no significant difference between the first and second order approximations. As a final remark, we note that an infinite order approximation would give the population dynamics result: in this case, the infinite temperature solution $\bar{\omega}$ corresponds to a particular initial condition for $\mathcal{P}$, which is lost after a large number of iterations of the approximation scheme. 

\begin{figure}[t]
\centering
\includegraphics{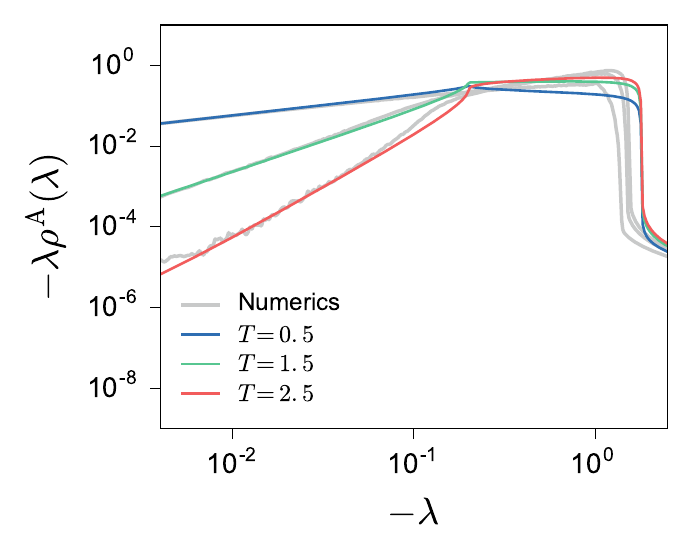}\includegraphics{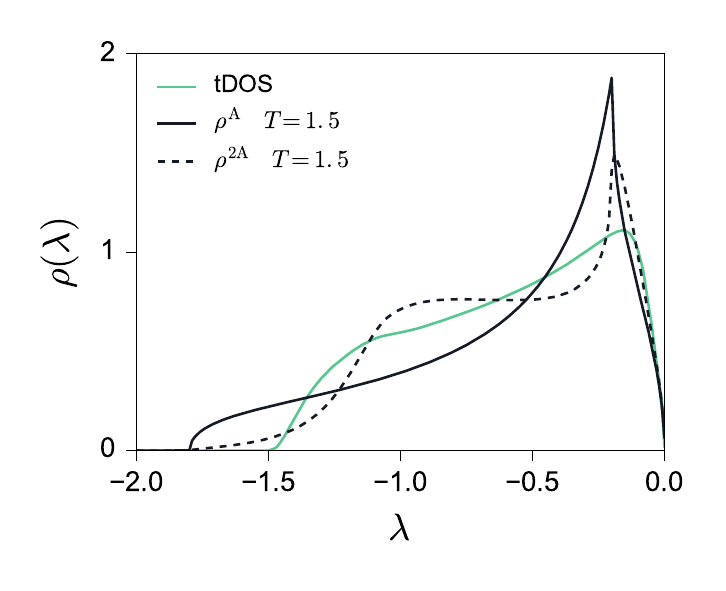}
\caption{Total DOS for mean connectivity $c=5$ and temperatures $T=0.5,1.5,2.5$. Left: first order approximation compared to spectra from direct diagonalisation (grey, statistics from $M=10^4$ system samples of size $N=1000$). Right: population dynamics prediction (green),  first order (black) and second order (dashed) approximations for $T=1.5$. All the evaluations have been performed using $\varepsilon \sim 10^{-4}$ and a population of size $N_{\mathrm{p}} = 2500$.}\label{fig:Approx}
\end{figure}

\subsection{Extended DOS and IPR}

\begin{figure}[t]
\centering
\includegraphics{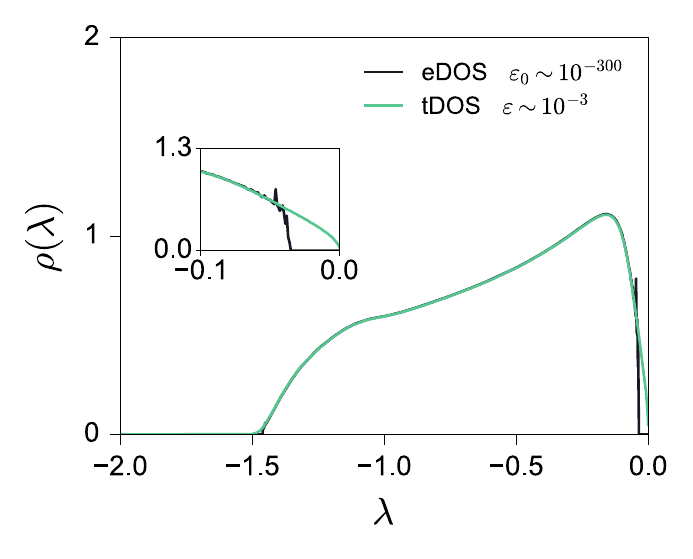}\includegraphics{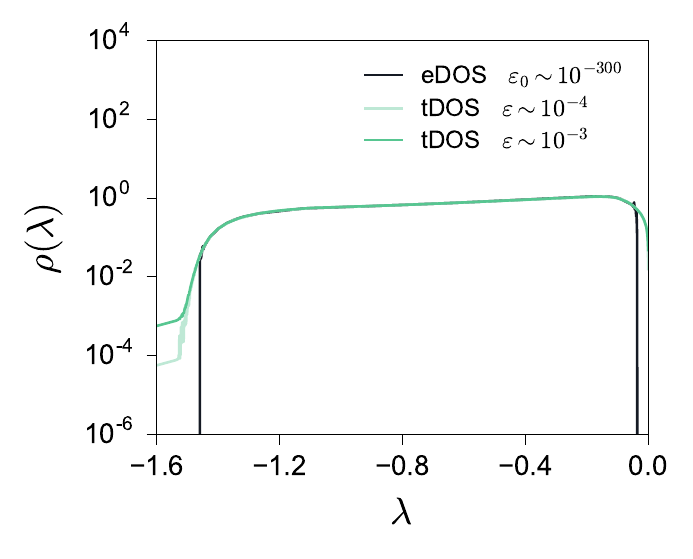}
\caption{Left: total DOS (green, tDOS) and extended DOS (black, eDOS) for connectivity $c=5$ and temperature $T=1.5$. Inset: zoom on the localisation transition occurring on the right edge of the spectrum; the mobility edge lies at $\lambda_c\simeq -0.04$. The evaluations have been performed using a population of size $N_{\mathrm{p}}=2500$. The noisy peaks in eDOS are due to the slow convergence of the algorithm at the localisation transition and should be ignored as unphysical. Right: same plot with a logarithmic $y$-scale, including total DOS evaluations with two different $\varepsilon$ values.}\label{fig:DOSlininset}
\end{figure}

As explained above, we can measure the extended DOS (only) by evaluating the cavity predictions in the limit $\varepsilon\to 0$ as it is this part of the spectrum that becomes continuous in the thermodynamic limit. Comparing the extended and total DOS then allows us to locate the mobility edges of the system. Figure \ref{fig:DOSlininset}-left shows the total DOS and the extended DOS on a linear scale, with an inset zooming in on the localisation transition occurring on the right end of the spectrum. Figure \ref{fig:DOSlininset}-right displays the same plot with a logarithmic $y$-axis, where we have included evaluations of the total DOS for different $\varepsilon$ values. This allows one to estimate the left end of the spectrum from the point on the $\lambda$-axis where the total DOS ceases to be $\varepsilon$-independent. Note that on approaching the mobility edges, the convergence of the population dynamics to its steady state becomes very slow. The  peaks in the extended DOS that are visible in the inset of figure \ref{fig:DOSlininset}-left are caused by this and should accordingly be ignored as unphysical. While it is not surprising to find localisation tails at the edges of the spectrum, at least from a random matrix perspective, it is remarkable that the appearance of mobility edges arises directly from the combination of two limiting cases with exclusively extended (RW) and localised (MF) eigenvectors, respectively. We can already argue that the fastest and slowest processes are governed by localised modes, with an intermediate regime where all modes are delocalised. Since we are interested in the long time dynamics, our attention will be focused on the bulk of extended states and on the slow (MF) localised modes only; we will also show that the fraction of fast localised states is relatively small compared to that of slow modes. As we will see, the mobility edge occurring on the slow end of the spectrum (of eigenvalues, or similarly relaxation rates) allows one to identify three different regimes in the time domain.

\begin{figure}[t]
\centering
\includegraphics{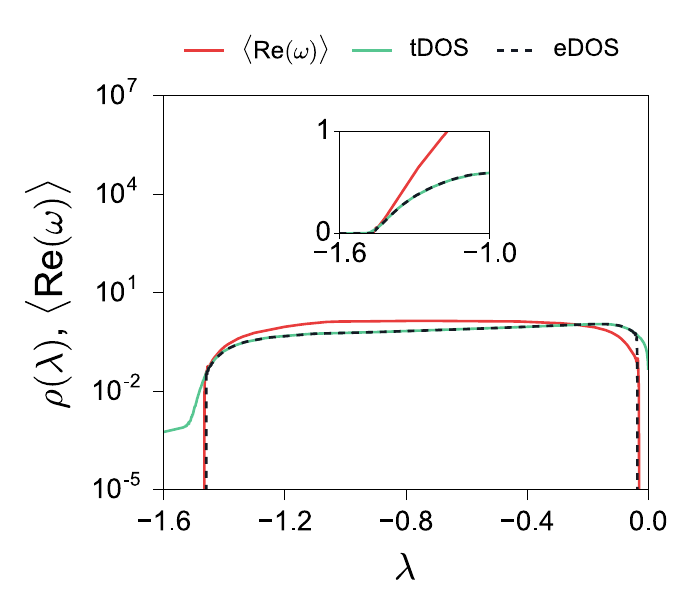}\includegraphics{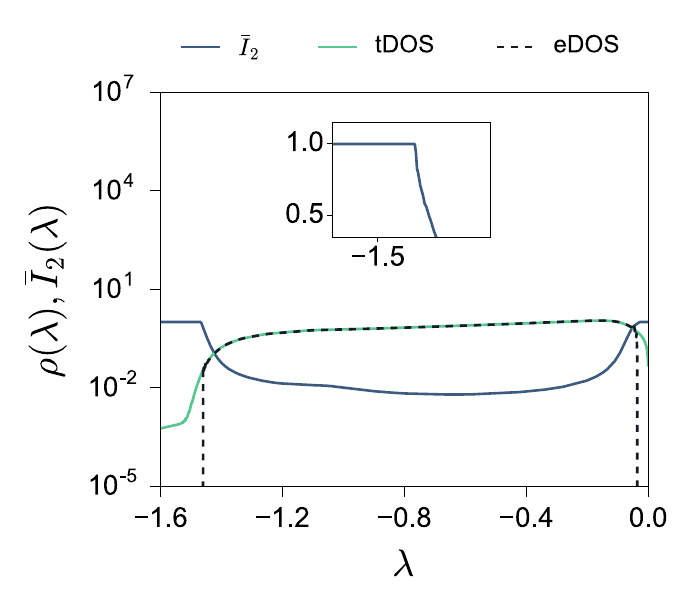}
\caption{Left: Average real part of the cavity precisions (red), extended DOS (black dashed line) and total DOS (green) for connectivity $c=5$ and temperature $T=1.5$. In the localised part of the spectrum, the cavity precisions have vanishing real part, i.e.\ are purely imaginary. The inset shows a zoom on the left localisation transition on a linear scale. The average real part of the precisions drops continuously to zero. Right: average IPR $\bar{I}_2(\lambda)$ (blue) predicted by (\ref{IPR}), alongside extended DOS (black) and total DOS (green dashed line). In the extended region of the spectrum the IPR scales with $\varepsilon$ as expected. The inset shows a zoom on the left localisation transition on a linear scale, where in the localised region $\bar{I}_2(\lambda)$ equals unity. The evaluations have been performed using a population of size $N_{\mathrm{p}}=2500$. For the total DOS and $\bar{I}_2(\lambda)$ we have used $\varepsilon \sim 10^{-3}$; for the extended DOS we have used $\varepsilon_0\sim 10^{-300}$. }\label{fig:ReOm}
\end{figure}

The localisation transition described above is associated with a change in the distribution $p(\omega)$: the population of cavity precisions converges to a steady state which has complex support for $\lambda$ in the bulk of the spectrum, and purely imaginary support outside. 
This transition can be detected by considering the average real part of the cavity precisions, which is shown in figure \ref{fig:ReOm}-left: 
as these precisions must have non-negative real part, the vanishing of the \emph{average} real part means all real parts are zero. ``Zero'' is to be interpreted here as of order $\varepsilon_0$, the value of $\varepsilon$ used in the population dynamics; our $\varepsilon_0  \sim 10^{-300}$ is indistinguishable from zero even on the logarithmic scale of figure \ref{fig:ReOm}-left. The figure shows the average real part of the cavity precisions and the total/extended DOS as a function of $\lambda$. As claimed above, the average real part is nonzero in the bulk of the spectrum, goes to zero exactly where the extended DOS does, and then vanishes within the localised spectrum. The approach of the average real part to zero is continuous (see inset), indicating that the transition in the structure of the distribution of cavity precisions is likewise continuous.

Note that having imaginary cavity precisions amounts to having real diagonal entries of the resolvent (via eq. (\ref{Resolvent_entries_precision})), which in the context of field theory and Anderson localisation are related to the so called self-energies. Similarly to what we have outlined above,
the state of an electron in a disordered medium is classified as localised or delocalised depending on whether the electron's self-energy is real or complex \cite{Abou-Chacra1973}.

We complement the above results by measuring the average degree of localisation of the eigenvectors.
For $N\to\infty$ one cannot access the IPR of individual eigenvectors. Instead one can consider the average IPR in a small range $\varepsilon$ around $\lambda$ and then take $\varepsilon$ to zero: 
\begin{equation}
\bar{I_2}(\lambda)=\lim_{\varepsilon\to 0}\lim_{N\to \infty}\frac{1}{N\rho(\lambda)}\sum_{\alpha=0}^{N-1}\delta_\varepsilon(\lambda-\lambda_{\alpha})I_2(\mathbf{v}_{\alpha})
\end{equation}
where $\delta_\varepsilon(x)=\varepsilon/[\pi(x^2+\varepsilon^2)]$ is a Lorentzian of width $\varepsilon$ and $\rho(\lambda)$ is assumed to be calculated similarly, using $\delta_\varepsilon(x)$ instead of $\delta(x)$ in the definition (\ref{DOS}). The order of the limits in the definition ensures self-averaging because the number of $\lambda_\alpha$ that contribute, which is of order $N\varepsilon$, becomes large.

Swapping the two limits, i.e.\ assuming that at any given $\lambda$ at most one eigenvector contributes to the average IPR, one can relate $\bar{I}_2(\lambda)$ to the squared modulus $|G_{jj}(\lambda)|^2$ of the resolvent entries. Boll\'e \emph{et al}.\ obtained from this a formula that allows the IPR to be evaluated within population dynamics \cite{Metz2010}, and used this to study localisation transitions in Laplacian and Levy matrices.  In our notation their expression reads
\begin{equation}\label{IPR}
\bar{I}_2(\lambda)=\lim_{\varepsilon\to 0}\frac{\varepsilon}{\pi\rho(\lambda)}\Big\langle \Big| \frac{\tau c}{\Omega_{c}(\{\omega_l\},\tau)} \Big|^2 \Big\rangle_{\{\omega_l\},\tau}
\end{equation}
which can be rewritten more explicitly as
\begin{equation}\label{IPR_A}
\bar{I}_2(\lambda)=\lim_{\varepsilon\to 0}\Big\langle \frac{\varepsilon }{(\varepsilon+A_{\mathrm{r}})^2+(\lambda+A_{\mathrm{i}})^2} \Big\rangle\Big/\Big\langle \frac{\varepsilon+A_{\mathrm{r}}}{(\varepsilon+A_{\mathrm{r}})^2+(\lambda+A_{\mathrm{i}})^2}\Big\rangle
\end{equation}
where, to keep the notation simple, we have used
\begin{equation}\label{A}
A = A_{\mathrm{r}} +\mathrm{i}A_{\mathrm{i}} = \frac{\Omega_c}{\tau c} - \mathrm{i}\lambda_{\varepsilon} = \frac{1}{\tau c} \sum_{l=1}^c\frac{\mathrm{i}\omega_l}{ \mathrm{i}+\omega_l}
\end{equation}
with $A_{\mathrm{r}}$ and $A_{\mathrm{i}}$ respectively the real and imaginary part of $A$. If the cavity precisions have zero/positive real part, then $A_{\mathrm{r}}$ is zero/positive accordingly. It follows from (\ref{IPR_A}) that $\bar{I}_2(\lambda)=1$ in the localised part of the spectrum, where the distribution $p(\omega)$ has purely imaginary support, and it is of order $\varepsilon$ within the bulk, where the support of $p(\omega)$ is complex.
While we expect an average IPR of \emph{order} unity, a value exactly equal to one is implausible in our case. This can be seen from the large $c$-limit, where we must recover the IPR of the MF eigenvectors (\ref{MF_eigenvector_element}), for which clearly $\bar{I}_2 < 1$ (see also figure \ref{fig:MF_RW}-right). The discrepancy indicates that the swapping of the limits $\varepsilon \to 0$  and $N\to \infty$ is not in general justified. 
Nonetheless, (\ref{IPR}) remains useful as a tool for differentiating between localised and extended parts of a spectrum.

As an alternative to the treatment of Boll\'e {\it et al}, we suggest an approximation to the IPR that is derived by taking $N\to\infty$ at fixed $\varepsilon$, and therefore is suitable for use within population dynamics based on (\ref{Zeta}). We leave the derivation to appendix \ref{appendixD} and only give the result
\begin{equation}\label{IPR2}
\bar{I}_2^{\star}(\lambda) = \lim_{\varepsilon\to 0} \frac{2\varepsilon}{\pi \rho(\lambda)}\text{Var}\Big[\text{Re}\Big(\frac{\tau c}{\Omega_c(\{\omega_l\},\tau)}\Big)\Big]_{\{\omega_l\},\tau}
\end{equation}
where $\text{Var}(\cdot)$ indicates the variance. The order of limits ($N\to\infty$ first, then $\varepsilon\to 0$) used ensures that there are always enough states within the $\lambda$-range of width $\varepsilon$ where quantities are measured. 
In this regard, our approach is opposite to that of Boll\'e \emph{et al}, where the two limits are inverted. Even so, we observe a very close agreement between the IPR estimates (\ref{IPR}) and (\ref{IPR2}), as shown in figure \ref{fig:IPRs} -- appendix \ref{appendixD}.

Figure \ref{fig:ReOm}-right shows the total DOS, the extended DOS, and the average IPR predicted by (\ref{IPR}). As explained above, in the localised region $\bar{I}_2(\lambda)$ has a constant value of one, even where the total DOS drops to the $\varepsilon$-value used in the measurement step of the population dynamics algorithm, i.e.\ outside of the support of the spectrum. The localisation transitions, detected as the points on the $\lambda$-axis where the value of $\bar{I}_2$ changes from $\mathcal{O}(\varepsilon)$ to $\mathcal{O}(1)$, occur where the extended DOS drops to $\mathcal{O}(\varepsilon_0)$. This confirms what we have discussed before: in the thermodynamic limit the spectrum has a continuous part of extended states, with localisation tails of pure point states occurring at the edges of the bulk.

\subsection{Finite size effects}

\begin{figure}[t]
	\centering
	\includegraphics{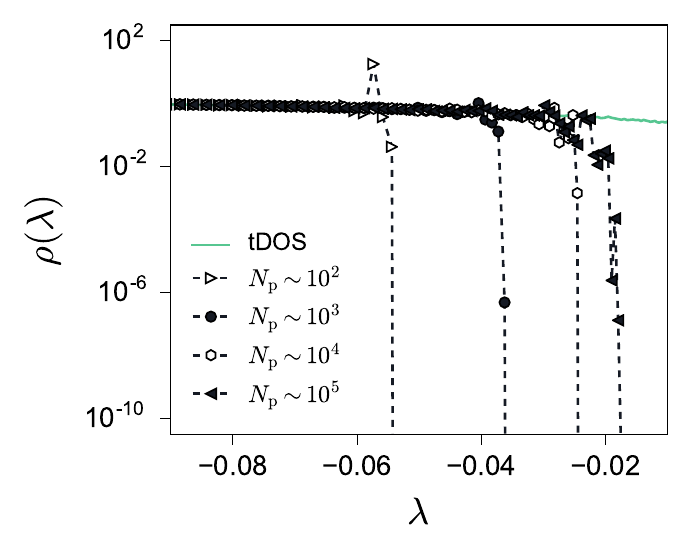}\includegraphics{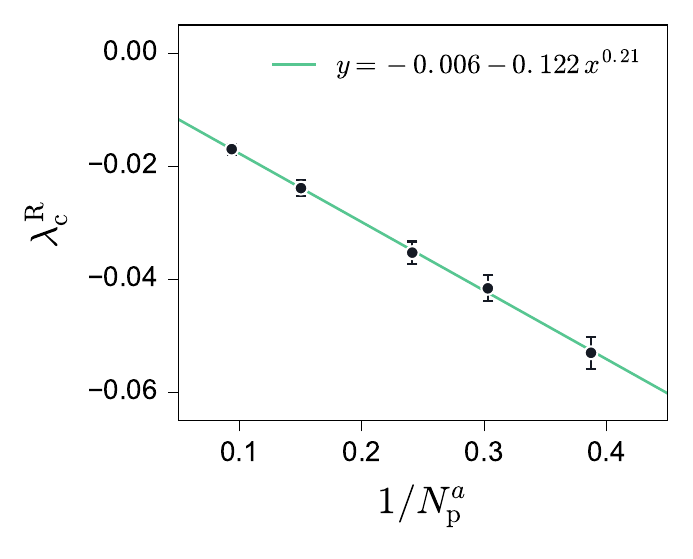}
	\caption{Left: extended DOS for different population sizes $N_{\mathrm{p}}$ (black) and total DOS (green), for connectivity $c=5$ and temperature $T=1.5$. Right: least squares fit of $\{(\lambda^{\mathrm{R}}_{\mathrm{c}}(N_{\mathrm{p}}),N_{\mathrm{p}}^{-a})\}$ giving $\lambda_{\mathrm{c}}\big|_{N_{\mathrm{p}}\to \infty} \simeq -0.006$.}\label{fig:Np_FSE}
\end{figure}

\begin{figure}[h]
	\centering
	\includegraphics{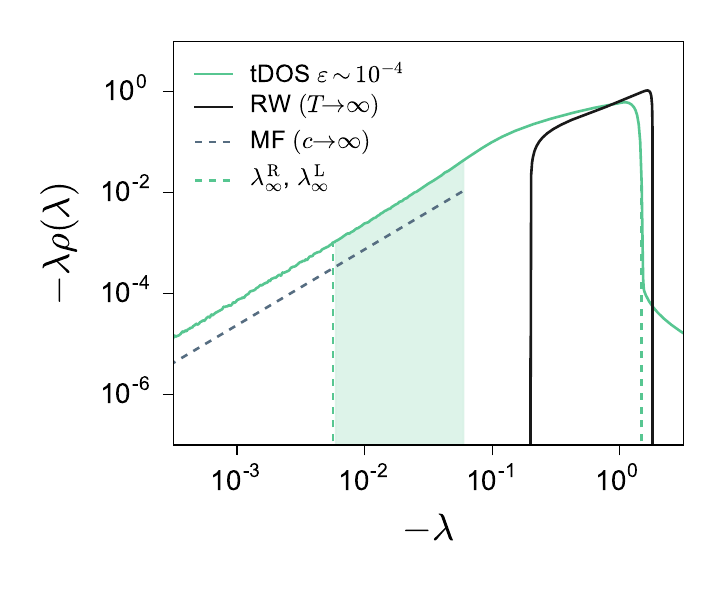}\includegraphics{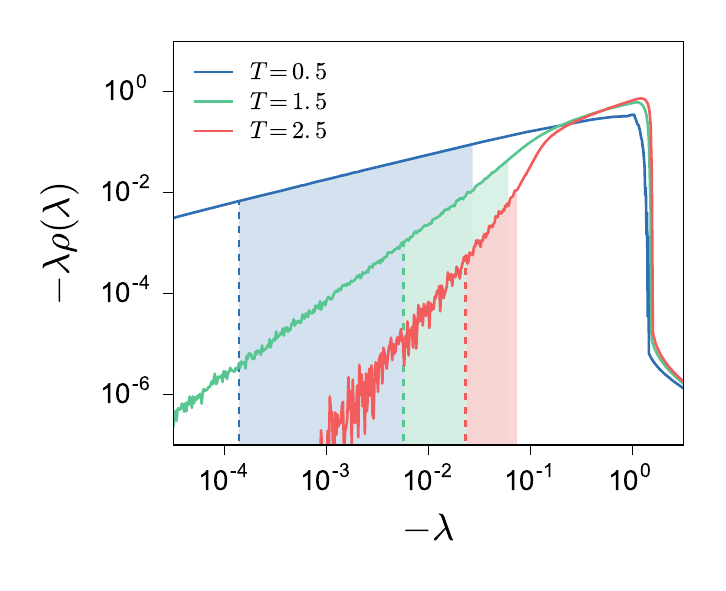}
	\caption{Left: total DOS (green solid line), extrapolated right and left mobility edges (green dashed line), occurring respectively on the left and right sides of the plot, MF DOS (power law, blue dashed line) and RW DOS (black solid line) for $c=5$ and $T=1.5$; the RW DOS is evaluated via (\ref{RW_DOS}). The spectrum is composed of three main regions: MF-localised (left), MF-extended (centre, green shaded area) and RW-extended (right). Right: total DOS (solid lines) and extrapolated right mobility edges (dashed lines) for $c=5$ and $T=0.5$ (blue), $1.5$ (green), $2.5$ (red). These spectra show the same qualitative features discussed for the case of $T=1.5$, but with the fraction of modes in the central ``mixed'' (MF-extended) region increasing as $T$ decreases. The evaluations of the total DOS have been performed using $\varepsilon\sim10^{-5}$ and a population of size $N_{\mathrm{p}}=2500$.}\label{fig:Loc_trans}
\end{figure}

\begin{figure}[htbp]
\centering
\includegraphics{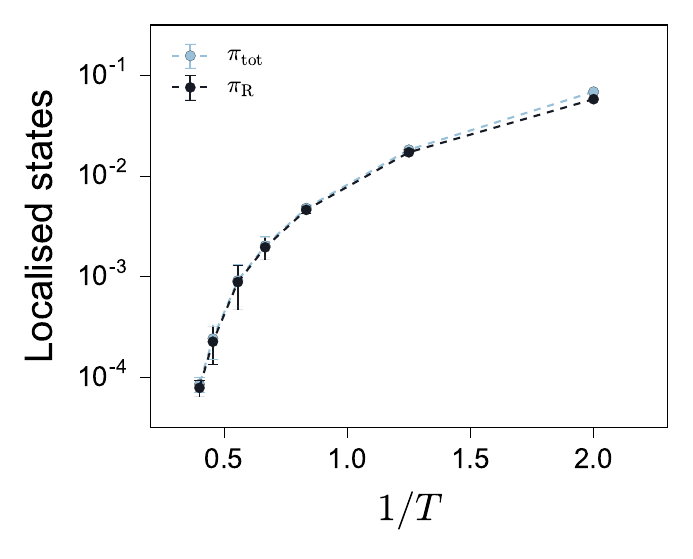}\includegraphics{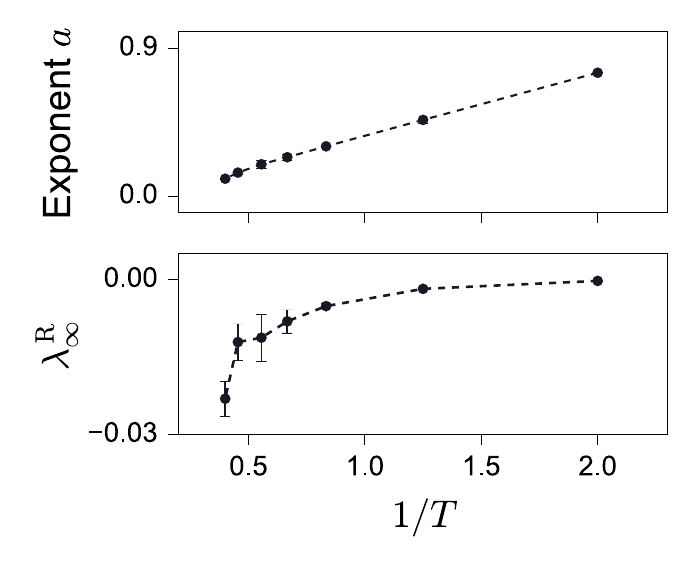}
\caption{Left: DOS integrated over the localised part of the spectrum to give the total fraction of localised modes $\pi_{\mathrm{tot}}$ and the fraction of slow localised modes $\pi_\mathrm{R}$ against inverse temperature; note that the two curves almost overlap. Right: exponent $a$ (top) and right mobility edge (bottom) extrapolated from the least squares fit (see equation (\ref{lam_c_N_scaling})), plotted against inverse temperature. These evaluations have been performed with mean connectivity $c=5$.}\label{fig:fractions_a_lambda}
\end{figure}

The advantage of the population dynamics approach is that it allows us to evaluate the spectral properties of infinitely large systems at a relatively low computational cost. As described above, the distribution $p(\omega)$ is approximated by a large population $\mathcal{P}$ of representative cavity precisions samples. This population converges to steady states that depend on the value of $\lambda$, and it undergoes critical transitions at the mobility edges. The location of these transitions turns out to have a non-negligible dependence on the population size $N_{\mathrm{p}}$ (see figure \ref{fig:Np_FSE}-left). Finite size effects in population dynamics algorithms have been discussed in the context of the Moran model \cite{Traulsen2005} and in the evaluation of large deviation functions \cite{Nemoto2016, Nemoto2017}. In particular in \cite{Nemoto2017} the authors show that in their case, systematic errors in the algorithm decrease proportionally to the inverse of the population size.
In order to determine the actual position of the mobility edges occurring within our spectra we assume that
\begin{equation}
|\lambda^{\mathrm{L/R}}_{\mathrm{\infty}}-\lambda^{\mathrm{L/R}}_{\mathrm{c}}(N_{\mathrm{p}})| \sim N_{\mathrm{p}}^{-a}
\end{equation}
where $\lambda^{\mathrm{L/R}}_{\mathrm{c}}(N_{\mathrm{p}})$ is the left/right mobility edge measured using a population of size $N_{\mathrm{p}}$, and $\lambda^{\mathrm{L/R}}_{\mathrm{\infty}}=\lim_{N_{\mathrm{p}}\to \infty}\lambda^{\mathrm{L/R}}_{\mathrm{c}}(N_{\mathrm{p}})$. Accordingly, we gather data for different $N_{\mathrm{p}}$ and  then fit them using 
\begin{equation}\label{lam_c_N_scaling}
\lambda^{\mathrm{L/R}}_{\mathrm{c}}(N_{\mathrm{p}}) = c_1 + c_2 N_{\mathrm{p}}^{-a}
\end{equation}
where $c_1$, $c_2$ and the exponent $a$ are determined by minimizing the least-squares deviation. This then identifies, in particular, the extrapolated value $\lambda^{\mathrm{L/R}}_{\mathrm{\infty}} =c_1$.
As shown in figure \ref{fig:Np_FSE}-right, with the exponent $a$ chosen in this way our data $\{(\lambda^{\mathrm{L/R}}_{\mathrm{c}}(N_{\mathrm{p}}),N_{\mathrm{p}}^{-a})\}$ do lie on a straight line to a good approximation as (\ref{lam_c_N_scaling}) assumes. This fitting method is applied to determine the right and left mobility edges for different values of the temperature, with mean connectivity $c=5$. 
The exponent $a$ is non-trivial in our case: it shows a monotonic increase with inverse temperature and typically lies between 0 and 1 (see figure \ref{fig:fractions_a_lambda}, top-right), in contrast to the setting in \cite{Nemoto2017} where $a=1$. 
We conjecture that the $T$-dependence of $a$ is related to the fact that also the exponent of the distribution of waiting times $\rho_{\tau}(\tau)$ varies with $T$; a more precise quantitative understanding of the value of $a$ remains an open problem, however.
Figure \ref{fig:Loc_trans}-left shows the DOS with the extrapolated right $\lambda_{\mathrm{\infty}}^{\mathrm{R}}$  and left $\lambda_{\mathrm{\infty}}^{\mathrm{L}}$ mobility edges (respectively on the left and right of the plot, because the $x$-axis shows $-\lambda$), the RW-DOS and the power-law MF-DOS for the slow modes. We note that $\lambda_{\mathrm{\infty}}^{\mathrm{R}}$ lies at a point on the $\lambda$-axis where the full DOS is already MF-like. In facts, it seems natural to describe the spectrum as composed of three main regions: the slowest modes possess MF-like features as they are localised and power-law distributed. The fastest modes are delocalised and exhibit a non-monotonic DOS that is closely related to  the Kesten-McKay law for the RW limit. Finally, the intermediate region (green shaded area) has mixed properties of the two limiting cases: here the eigenstates are delocalised (RW-like) but show a power-law distribution (MF-like). We also observe that this intermediate region becomes wider -- the fraction of delocalised modes with a MF-like density of states increases -- as the temperature decreases (figure \ref{fig:Loc_trans}-right).
 
From the extrapolated position of the mobility edges in the eigenvalue spectrum we can estimate the fraction of localised modes $\pi_{\mathrm{tot}}$ in our system. This is given by the integral of the total DOS over the $\lambda$-regions containing localised states, which in our case lie at the edges of the spectrum. It is in fact simpler to evaluate $\pi_{\mathrm{tot}}$ by working out the complement, i.e.\ integrating over the bulk of the DOS:
\begin{equation}
\pi_{\mathrm{tot}}=1-\int_{\lambda_{\mathrm{\infty}}^{\mathrm{L}}}^{\lambda_{\mathrm{\infty}}^{\mathrm{R}}} \mathrm{d}\lambda\, \rho(\lambda)
\end{equation}
Here $\lambda_{\mathrm{\infty}}^{\mathrm{L}}$ and $\lambda_{\mathrm{\infty}}^{\mathrm{R}}$ are the left and right mobility edges of the system, extrapolated to infinite population size as explained above. 

We can similarly obtain the fraction of localised fast/slow modes by integrating over the $\lambda$-region at the left/right end of the spectrum. Since we have Lorentzian tails of width $\varepsilon$ affecting the total DOS, the most accurate way of computing these fractions is to locate the left end of the spectrum $\lambda_{\mathrm{end}}^{\mathrm{L}}$ by exploiting the $\varepsilon$ dependence of the total DOS (see figure \ref{fig:DOSlininset}-right), then evaluating the fraction of localised fast modes as 
\begin{equation}
\pi_{\mathrm{L}} = \int_{\lambda_{\mathrm{end}}^\mathrm{{L}}}^{\lambda_{\mathrm{\infty}}^{\mathrm{L}}} \mathrm{d}\lambda\, \rho(\lambda)
\end{equation}
The fraction of localised slow modes is finally given by $\pi_{\mathrm{R}} = \pi_{\mathrm{tot}} - \pi_{\mathrm{L}}$. Figure \ref{fig:fractions_a_lambda} shows the fractions of localised modes (left) and the right mobility edge (bottom-right) as functions of the temperature, for mean connectivity $c=5$. We observe that as the temperature decreases the localisation region on the right edge of the spectrum becomes narrower while the fraction of slow localised modes in this increases. Overall, the total fraction of localised eigenstates becomes larger as the temperature decreases. Nevertheless, even at $T=0.5$ -- the lowest temperature considered here -- the fraction of localised modes only amounts to around $10$\% of the total DOS. The majority of these localised eigenstates lie in the low $|\lambda|$ tail, as can be seen from the fact that the quantities $\pi_{\mathrm{R}}$ and $\pi_{\mathrm{tot}}$ are almost overlapping on the log scale shown in figure \ref{fig:fractions_a_lambda}-left. Importantly for the long-time dynamics, all the slowest modes in the system are localised, at least for the temperature regime that we have considered here.
The temperature trend for low $T$ is consistent, at the other end, with the $T\to\infty$ limit: here we obtain a RW spectrum with only extended and no localised modes. 

\section{More disordered network topologies}\label{section5}

So far we have focused on the case of random regular graph (RRG) connectivity, where the network defining the possible paths among minima in the potential energy landscape has a regular structure that becomes free of disorder in the thermodynamic limit. From a topological perspective the absence of disorder might seem as unrealistic as, say, in the $n$-dimensional hypercubic lattice or the complete graph with its mean field connectivity. However, the RRG does introduce the essential features of sparse random networks, i.e.\ it is ``infinite dimensional'' -- the number of nodes grows exponentially with distance -- and it confines all dynamical transitions to a local environment. The RRG case is also interesting as the localisation properties of the eigenvectors of the master operator are the opposite of those in the mean-field Bouchaud model, where the connectivity is infinite and all eigenmodes are power-law localised on the energy axis. 

The question that we want to address in this section is whether the RRG case possesses all the relevant features of sparsely connected energy landscapes, at least in terms of the spectral properties discussed so far, or whether more disordered network topologies add new features (see e.g. \cite{Regev2017}). We therefore extend our analysis to  Erd\"os-R\'enyi (ER) and scale-free (SF) graph structures, which have been widely studied in other contexts \cite{Albert2002, Lu2006}. 
They both have finite average degree $c$ but are paradigmatic as graph ensembles with finite (ER) and infinite (SF) degree variance, respectively.
Further motivation comes from the fact that numerical studies on a relatively small number of Lennard-Jones interacting atoms have suggested a configuration-space connectivity of the scale-free type \cite{Doye2002}, which is also the network topology assumed by Baronchelli \emph{et al} \cite{Moretti2011}
as discussed in the introduction. The SF case may therefore represent the best candidate for modelling configuration space connectivity, though we stress that our approach is flexible and can be applied to any network topology without short loops. 

Looking at the random walk (RW) and mean-field (MF) limits, we note first that in the former case there is no simple closed form expression for the DOS, analogous to (\ref{RW_DOS}) for regular graphs, on complex network structures: we will have to obtain results by population dynamics instead for the $T\to\infty$ limit.
The limit $c\to\infty$, on the other hand, effectively brings us back to the fully connected case so our previous results and discussion for the MF limit still apply.

Erd\"os-R\'enyi graphs \cite{Erdos1959} of size $N$ are constructed by assigning an edge between any pair of vertices with probability $p$, so the average number of edges in the network is $N(N-1)p/2$ and we need $(N-1)p$ to be finite to ensure that the resulting graphs are sparse. The probability that a given node has $k$ neighbours then follows a binomial distribution, which in the large $N$ limit approaches a Poisson distribution with parameter $c=(N-1)p$. We therefore apply our cavity method assuming $p_k = e^{-c} c^k/k!$, with $\langle k \rangle = \text{Var}(k)=c$. Since the Poisson distribution is strongly peaked around $c$, the local environment of these graphs is typically subject to weak fluctuations, and the overall structure is not far from that of random regular graphs.
In the following we will assume $c=5$, which ensures that the fraction of nodes in the giant cluster is approximately equal to one, ignoring the effects of very small disconnected components on the spectral properties of the whole system. The population dynamics algorithm applies as explained in section \ref{section4}, with the only difference that at each update we pick $k-1$ elements from the population of cavity marginals with probability $p_k k/c$; see appendix \ref{appendixB} for further details.

\begin{figure}[h]
	\centering
	\includegraphics{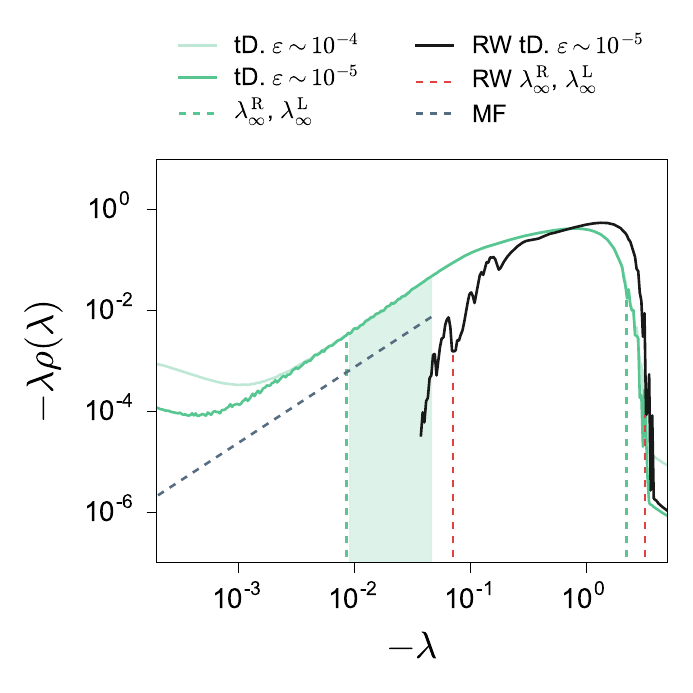}\includegraphics{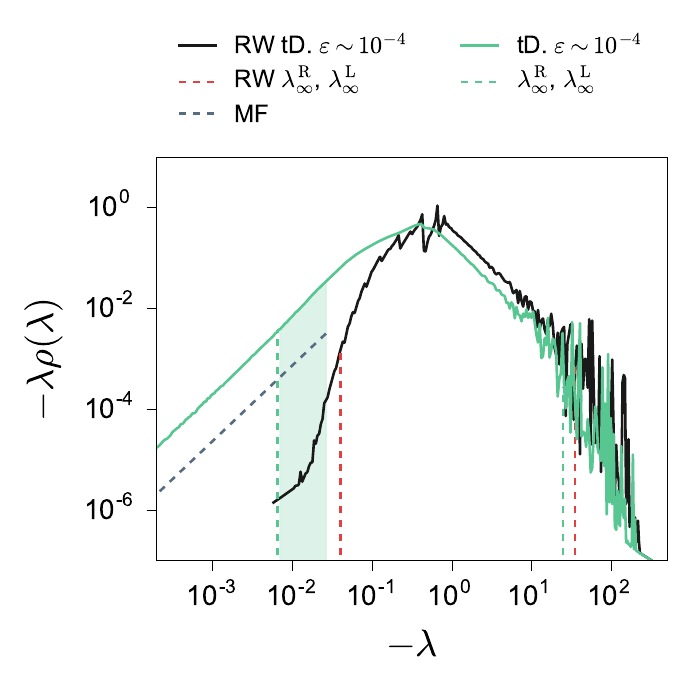}
	\caption{Left: spectral attributes for Erd\"os-R\'enyi networks. Total DOS (green solid lines) and extrapolated mobility edges (green dashed lines) for $c=5$ and $T=1.5$, with the corresponding MF DOS (power law, blue dashed line), RW DOS (black solid line) and extrapolated RW mobility edges (red dashed lines). The spectrum is composed of three main regions: MF-localised (left), MF-extended (centre, green shaded area) and RW-like (right). Right: analogous plot for scale-free networks with system parameters $\gamma = 2.5$ (implying $c\simeq 4.53$), and $T=1.5$. The same colour scheme of the left-hand plot applies. Both plots have $-\lambda$ on the $x$-axis, making the right mobility edges appear on the left side of the pots (and vice-versa for the left mobility edges).  The quantities tDOS, RW eDOS and RW tDOS in both plots have been computed using a population of size $N_{\mathrm{p}}=2500$.}\label{fig:ER_SF}
\end{figure}

The spectral features of the ER ensemble with $c=5$ and $T=1.5$ are shown in figure \ref{fig:ER_SF}-left. The total DOS is displayed for evaluations involving two different values of $\varepsilon$, whose effect is visible on the left of the plot. The total DOS of the RW limit would have the same $\varepsilon$ tail for small $|\lambda|$ but we do not show this region as the RW DOS becomes too small to estimate reliably there. Similarly to the case of random regular graphs, the DOS is composed of three main parts: a mean field power-law tail occurs at the slow end of the spectrum, covering localised (left) and delocalised (centre) modes, while the distribution of fast modes (right) is non-monotonic and follows closely the DOS of the corresponding RW limit. In contrast to the RRG case, the connectivity disorder alone is enough to induce localisation transitions within the spectrum; the mobility edges are extrapolated by the least squares fit discussed in the previous section, and they are marked by the green ($T=1.5$) and red ($T\to\infty$) dashed lines in the plot. We observe that the area under the total DOS of MF localised modes at $T=1.5$ is much larger than that corresponding to the RW case. This means that the intrinsic localisation attributes of ER graphs are sub-dominant with respect to the effects introduced by energy disorder, which become stronger when the temperature is lowered.

The last class of networks that we address in this work is the scale-free type. In the form originally proposed \cite{Barabasi1999}, these networks are constructed via preferential attachment: starting with a dimer of two nodes linked together, one connects a new node to the existing ones with a probability that is proportional to the number of links that they already have, repeating the process until the network has the desired size. More generally SF networks are characterised by a degree distribution $p_k$ that decays as $k^{-\gamma}$. Here $\gamma$ is typically in the range $2<\gamma<3$, implying that second and higher order moments diverge. This motivates the appellative ``scale-free" as, by contrast to Erd\"os-R\'enyi and random regular graphs, the degree fluctuations are infinitely large and have no intrinsic scale. Defining a SF graph ensemble by assigning equal probability to all networks with the given degree distribution, one typically finds many hubs within the network, i.e.\ nodes with very high degree, occurring at all (degree) scales. In order to minimise the number of disconnected sub-graphs we introduce a lower bound on the range of degrees by imposing $p_0 = p_1 = 0$. Also, in practice we cannot deal numerically with unbounded probability distributions, and a cutoff $K_{\mathrm{MAX}}$ has to be specified so that $p_k = 0 \,\,\forall\,\, k> K_{\mathrm{MAX}}$; we take $K_{\mathrm{MAX}}=1000$. The cutoff is entirely immaterial for the ER case, where for the Poisson degree distribution with e.g. $c=5$ one has $p_{K_{\mathrm{MAX}}} \sim 5^{1000}/1000! \sim 10^{-1869}$. Even for SF graphs, with $\gamma=2.5$, $p_{K_{\mathrm{MAX}}}\sim 1000^{-2.5}\sim 10^{-8}$ so the cutoff lies far in the tail of the distribution. It does make all moments of the distribution finite, but still retains much larger degree fluctuations than for random regular and ER graphs.

The results for the SF ensemble with $\gamma=2.5$ and $T=1.5$ are shown in figure \ref{fig:ER_SF}-right, using the same colour scheme as for the ER plot on the left and a population of $2500$ cavity precisions. A striking difference is that the spectrum is much broader than for the previous cases, by at least one order of magnitude (scaling the rates by the average connectivity is ineffective when the variance of degrees is large as here): a long tail of fast, localised modes appears (at the left end of the spectrum, which on the plot is on the right as the $x$-axis shows $-\lambda$). Similarly to the case of ER graphs, the area under the total DOS of MF localised modes at finite temperature ($T=1.5$) is much larger than that corresponding to the RW case, which demonstrates the strengthening of slow mode localisation when the temperature is lowered. Overall, in spite of some differences in the details, the DOS for SF networks has the same structure as for random regular and Erd\"os-R\'enyi graphs, with a tail of slow modes following the mean field statistics, a mixed region where the DOS is MF-like but eigenstates are delocalised, and a remaining part of the spectrum that is non-monotonic and closely related to the associated RW case. By lowering the temperature we induce a shift in the DOS towards slower modes, the range of MF-RW mixed modes becomes wider, and the fraction of slow localised modes increases.

\section{Conclusions and future perspectives}\label{section6}

In this paper we have considered the problem of walks on the potential energy landscape as described by the trap model of Bouchaud and others, extending previous analyses to the case of sparse inter-trap connectivity. In this scenario there are two different sources of disorder: one is associated with the topology defining the connectivity among minima, and the other one is given by the different energy depth of the traps. Accordingly there are two important notions of distance: the distance on the graph structure and the distance on the energy axis. The sparse structure of the master operator $\mathbf{M}$ makes the problem impossible to solve with analytical tools, and we then approached it by means of the cavity method, which in the thermodynamic limit leads to a population dynamics algorithm. This allowed us to evaluate the eigenvalue spectrum of $\mathbf{M}$, and the localisation properties of the associated eigenstates (the modes of the dynamics), which are key to understanding the dynamical behaviour of the model.

We first discussed the spectral properties of the ground state, i.e.\ the equilibrium distribution, focussing on how the IPR scales with system size for different temperatures; these results are independent of network structure because the transition rates obey detailed balance. 
In the bulk of the paper we considered the case of random networks with regular connectivity, where the key system parameters are the temperature $T$ and the mean connectivity $c$. We discussed the limiting situation of infinite temperature, where the dynamics is a random walk (RW) and only the distance on the graph structure matters. Here the density of states (DOS) is given by a shifted and scaled Kesten-McKay law, and all eigenmodes are delocalised. In the opposite mean-field (MF) limit $c\to\infty$, where traps are distinguished only by their energy depth, the eigenstates are localised and the DOS 
follows a power law with a $T$-dependent exponent. We found that these features are combined in the general case of finite connectivity and finite temperature where both notions of distance are relevant: a MF-like tail of slow localised modes (governing the long time dynamics) appears, while fast modes follow the RW case, being delocalised and showing a non-monotonic DOS related to the Kesten-McKay law.
Localisation transitions appear within the spectrum at the changeover between these two behaviours; they correspond to continuous transitions in the nature of the support of $p(\omega)$, the distribution of cavity precisions that is the key quantity within the population dynamics algorithm. The location of these transitions is affected by population size, and we extrapolated them to the infinite population limit using a simple power law form.
This revealed a surprise: the combination of RW and MF features give rise to a mixed region separating the fast modes from the slow modes, where the DOS has a MF shape but eigenstates are nonetheless delocalised. 

The shape of the DOS, and particularly the power law MF-like tail of slow modes, are well captured by a simple ``high temperature" approximation scheme. At first order, one cavity iteration (involving disorder) is performed starting from the infinite temperature solution found on the cavity graph (where, in the case of RRG, disorder is absent). This is similar to the ``single defect approximation" \cite{Biroli1999}: the central vertex is the only source of randomness and this allows one to find an analytical solution for the DOS.

We observed the same overall structure in the spectra for more disordered network topologies, specifically Erd\"os-R\'enyi and scale-free graphs, though with some changes in the details particularly for the fastest modes. The broader degree distributions of these networks are sufficient to induce localisation transitions even for infinite temperature, i.e.\ without any effects from the trap depths.
However, the fraction of slow localised modes is greatly enhanced at finite temperature, where the corresponding DOS has a power-law MF shape. The latter feature arises in every graph ensemble considered here. Bearing in mind that the spectrum of relaxation times is simply given by the collection of inverse eigenvalues $\{1/|\lambda_{\alpha}|\}$, the ultimate long time dynamics should then always be of mean field kind. This asymptotic independence from the network structure is in fact consistent with the way the trap model was originally designed, 
namely to describe dynamics in configuration-space on a very long time scale where deep minima are effectively fully connected by paths passing through shallow traps.
The presence of a region in the spectrum with mixed MF and RW properties suggests the existence of an intermediate time scale during which the dynamics should exhibit features of both the short-time random walk behaviour, which is network dependent, and the long time mean field evolution. How this distinction based on the spectral analysis can be quantified in the time domain remains an open problem and will have to be addressed in future. 

More generally, the lack of detailed information about the eigenvectors remains the major limitation of our approach, as it impedes a direct evaluation and classification of the ageing dynamics. Nevertheless, we believe that there is scope here for significant improvement and we hope that this work stimulates further investigations in this direction. In particular, these should include exploring the time domain and trying to characterise the three different regimes that we have highlighted above. This can be done e.g.\ by looking at the time-dependent probability of return to a given trap, which can be expressed as the Laplace transform of the local density of states (i.e.\ the contribution to the total DOS from a local node with a given trap depth). The local DOS will typically be peaked around the value of $-\lambda$ that corresponds to the initial decay of the return probability, and this would allow one to focus on a single one of the three distinct regions that we have identified in the spectrum. Alongside the return probability, other time dependent quantities, such as the mean number of distinct nodes visited within some time $t$, will help to characterise the (non-equilibrium) dynamics and identify the system's time scales. This question can be tackled with the techniques of \cite{Bacco2015}, accompanied by numerical simulations of random walks on trapping networks to assess finite size and pre-asymptotic (short $t$) effects.

Future work should aim also to elucidate the link between the dynamics and the localisation properties of glassy dynamics on networks. Insights might come from a closer look at the structure of the eigenvectors. The IPR carries no information on the spatial distribution of the eigenvector components, nor is it able to distinguish between exponential or non-exponential localisation, either in energy or on the graph structure. For this reason it is not clear from our results how the two sources of randomness influence the localisation strength of, say, the localised eigenstates governing the long time dynamics. The region of the spectrum that we have characterised as ``MF-localised'' might exhibit eigenvectors that are e.g.\ exponentially localised within small areas of the network, rather than covering all the nodes through a power law decay with difference in trap energy (as happens for mean field connectivity). Then, for waiting times in the range of the slow localised modes (which will be the typical case at low temperature), when the system escapes from a deep trap all delocalised modes will have decayed and the motion will remain confined to the neighbourhood of the initial node, in contrast to MF dynamics. Similarly, in the ``mixed'' region of the spectrum the delocalised eigenvectors might have non-trivial spatial structure; one might conjecture that they should be concentrated onto an extensive number of clusters of nodes, interpolating between the localised slow modes and the delocalised fast ones. Overall, therefore, more detailed information on the distribution of the entries of the slow eigenvectors and their spatial correlations should allow a better understanding of the asymptotic dynamics. The former quantity can easily be obtained numerically via e.g.\ the multifractality spectrum \cite{Fyodorov2010}, while the latter could be assessed with an analogue of the radial distribution function (or similar measures) from liquid state theory \cite{Hansen2013}. To calculate them in the large system size limit, on the other hand, as outputs from a population dynamics algorithm, remains a technical challenge.

Finally, we point out two possible extensions of the present work: given the evidence for correlations between the depth and the number of neighbours of an energy minimum, this would be a feature worth including within our model. Such a more general setting should be amenable to an analysis similar to the one in this paper as we sketch briefly in appendix \ref{appendixB}. In principle, correlations in the degree of neighbouring minima could also be taken into account. The idea, supported by previous works on simulations of L-J interacting atoms \cite{Doye2002}, is that deep minima in the potential energy landscape are surrounded by many shallower ones, creating a hierarchical structure and thus inducing degree-degree and degree-energy correlations.

The second interesting model extension would be to consider alternative transition mechanisms among minima, as considered e.g.\ by Barrat and M\'ezard \cite{Barrat1995, Bertin2003a}, who used Glauber transition rates. These rates depend on the difference in energy between the departure and arrival nodes, and as energy-decreasing transitions are always allowed one can picture the situation on a fully-connected graph as an energy landscape made up of steps rather than traps. Here the {\em entropy} of relaxation paths is key and leads to a completely different phenomenology (see for example \cite{Sollich2003, Sollich2006,Cammarota2015}), with ageing arising from entropic rather than energetic barriers. On sparse graphs, on the other hand, even Glauber dynamics will encounter energy barriers -- consider a deep trap with only shallow neighbours -- and so a much richer and possibly more realistic dynamics should result. Work towards analysing this case is in progress. 

%////////////////////////////////////////////////////////////////////////////////////////%

\section{Acknowledgements}\label{section7}

The authors acknowledge funding by the Engineering
and Physical Sciences Research Council (EPSRC)
through the Centre for Doctoral Training ``Cross Disciplinary
Approaches to Non-Equilibrium Systems''
(CANES, Grant Nr.\ EP/L015854/1). RGM gratefully acknowledges insightful discussions with Chiara Cammarota, Davide Facoetti and Aldo Glielmo.

%%////////////////////////////////////////////////////////////////////////////////////////%%
%%////////////////////////////////////////////////////////////////////////////////////////%%
%%                                                                                                                                                          %%
%%                                                                 Bibliography                                                                      %%
%%                                                                                                                                                          %%
%%////////////////////////////////////////////////////////////////////////////////////////%%
%%////////////////////////////////////////////////////////////////////////////////////////%%

\bibliographystyle{unsrt} % This gives abbreviated citations, sorted by order of appearance 
\bibliography{Paper_MKS} % Use the Bibliography.bib file as the source of references

%%////////////////////////////////////////////////////////////////////////////////////////%%
%%////////////////////////////////////////////////////////////////////////////////////////%%
%%                                                                                                                                                          %%
%%                                                                   Appendix                                                                        %%
%%                                                                                                                                                          %%
%%////////////////////////////////////////////////////////////////////////////////////////%%
%%////////////////////////////////////////////////////////////////////////////////////////%%

\appendix

\section{Localisation in mean field limit}\label{appendixA}

We give here a qualitative argument why in the mean-field limit a generic eigenvector associated with eigenvalue $\lambda<0$ will be localised. We regard $\lambda<0$ as fixed here and take $N\to\infty$, to stay well away from the groundstate. For finite $N$ one would expect a crossover to the localisation properties of the (delocalised, for $T>1$) ground state as $\lambda \to 0$.

The explicit form of the eigenvector components is given in (\ref{MF_eigenvector_element}). To estimate how these components vary across nodes $i$, consider a typical realisation of the trap depths $\{E_1, E_2, \ldots, E_N\}$, arranged in ascending order such that $E_i<E_{i+1}$. The inverse trapping times $\tau_i^{-1}=\exp(-\beta E_i)$, which determine the eigenvector components $u_i \propto (\lambda + \tau_i^{-1})^{-1}$ (we drop the eigenvector label $\alpha$ here), are then in descending order. The largest component will occur at the node $i$ with $\tau_i^{-1}$ closest to $|\lambda|$; call this node $j$. The number of inverse trapping times at other nodes that lie in an interval $[\tau_j^{-1},\tau_i^{-1}]$ is typically $N\rho_{\tau^{-1}}(\tau_j^{-1})
(\tau_i^{-1} - \tau_j^{-1})$ where $\rho_{\tau^{-1}}$ denotes the distribution of inverse trapping times. Abbreviating this density of states-factor as simply $\rho$, we can therefore write
\begin{equation}
j-i=N\rho(\tau_i^{-1}-\tau_j^{-1}) 
\end{equation}
as a deterministic approximation for the values of the inverse trapping times around $\tau_j^{-1}$.

Now call $S_{\mathrm{N}}$ and $S_{\mathrm{D}}$ respectively the sum on the numerator and denominator in the definition (\ref{IPR_q}) of $I_2$. Using that $\lambda \approx -\tau_j^{-1}$, we have
\begin{equation}
S_{\mathrm{N}}=\sum_i u_{i}^4\propto \sum_i (\lambda+\tau_i^{-1})^{-4} \simeq \sum_i \Big(\frac{j-i}{N \rho}\Big)^{-4}
\end{equation}
The last sum can be approximated as twice the integral over the positive values of $m=i-j$ 
\begin{equation}
S_{\mathrm{N}} \simeq 2 \rho^4 \int_{1}^{\infty}\mathrm{d}m\,\frac{1}{(m/N)^4}\propto \rho^4 N^4
\end{equation}
Similarly, for the sum in the denominator in $I_2$ we get
\begin{equation}
S_{\mathrm{D}} \simeq \Big( 2 \rho^2 \int_{1}^{\infty}\mathrm{d}m\,\frac{1}{(m/N)^2}\Big)^2\propto \rho^4 N^4
\end{equation}
Taking the ratio, it follows that $I_2 = O(1)$ for any eigenvector $\mathbf{u}$ with an eigenvalue away from zero; this statement holds at any temperature. Note that the sums or integrals defining $S_{\mathrm{N}}$ and $S_{\mathrm{D}}$ all converge at the upper end, i.e.\ have their mass concentrated around small $m=i-j$. This justifies our initial approximation of focussing on inverse trapping times close to $\tau_j^{-1}$. It also implies that the above argument for the IPR of the right eigenvectors of the master operator applies equally to the eigenvectors of the symmetric master operator: these differ only by factors of $\tau_i^{-1/2}$, which vary weakly (by $O(1/N)$) across the relevant range where $m=i-j$ is finite.
%////////////////////////////////////////////////////////////////////////////////////////%

\section{The cavity method}\label{appendixB}
\begin{figure}[htbp]
	\centering
	\includegraphics[height=2.5cm]{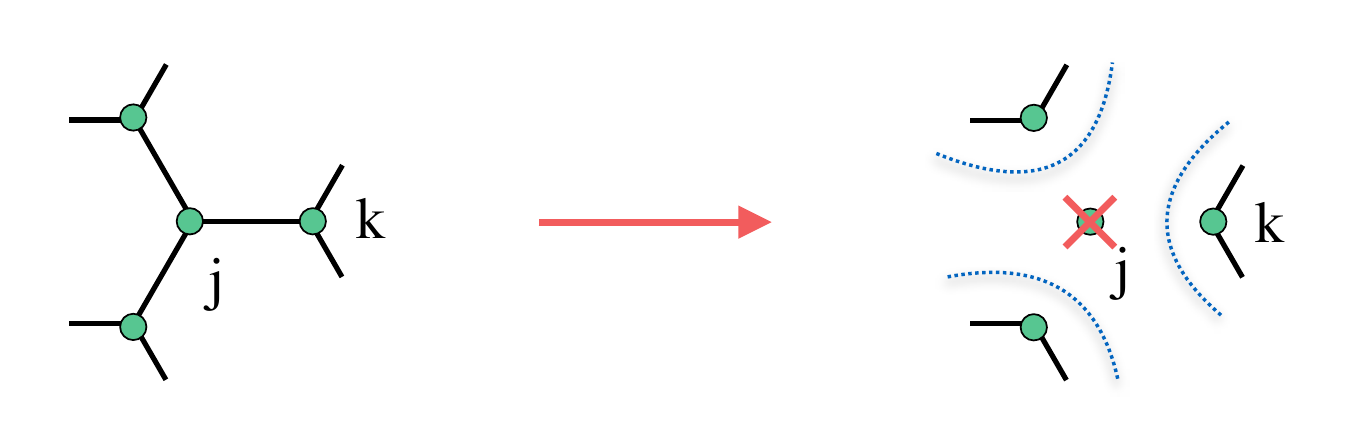}
	\caption{Neighbourhood of site $j$ on $\mathcal{G}$ (left) and on $\mathcal{G}^{(j)}$ (right). The red cross indicates that $j$ is absent in $\mathcal{G}^{(j)}$ and the branches become independent when the local structure is treelike.}\label{fig:Cavity_Graph}
\end{figure}

In this appendix we illustrate how to derive the equations (\ref{Precisions}) relating the (cavity) precisions, starting from the cavity marginal probability distributions as expressed in (\ref{Magrinal_cavity_distribution_y}), i.e. \
\begin{equation}\label{eq:B1}
P^{(j)}(y_k)=e^{-\frac{\mathrm{i}}{2}\lambda_{\varepsilon}y_k^2/r_j} \prod_{l\in \partial k\setminus j}\int  \mathrm{d}y_l \, e^{-\frac{\mathrm{i}}{2}(y_l-y_k)^2}P^{(k)}(y_l)
\end{equation}
which is based on the factorisation  (\ref{Cavity_assumption}).
As depicted in figure \ref{fig:Cavity_Graph}, this assumption works well if the graph lacks short loops, i.e.\ when it is locally treelike. Under this condition the correlations between the variables $\mathbf{y}_{\partial j}$ belonging to different branches become negligible when the common root is removed from the graph. Inserting the ansatz given in (\ref{Marginal_ansatz_distribution_y}), i.e.
\begin{equation}\label{eq:B2}
P^{(j)}(y_k)=\sqrt{\frac{\omega_k^{(j)}}{2\pi}}e^{-\frac{1}{2}\omega_k^{(j)}y_k^2}
\end{equation}
we obtain for the cavity marginals
\begin{equation}\label{eq:B3}
P^{(j)}(y_k)\propto e^{-\frac{\mathrm{i}}{2}\lambda_{\varepsilon}y_k^2/r_k} \prod_{l\in \partial k\setminus j}\int \mathrm{d}y_l\, e^{-\frac{\mathrm{i}}{2}(y_l-y_k)^2 - \frac{1}{2}\omega^{(k)}_l y_l^2}
\end{equation}
Completing the square and integrating out $y_l$ one finds
\begin{equation}\label{eq:B4}
\begin{split}
P^{(j)}(y_k) & \propto e^{-\frac{\mathrm{i}}{2}\lambda_{\varepsilon}y_k^2/r_k} \prod_{l\in \partial k\setminus j} e^{-\frac{\mathrm{i}}{2}y_k^2 - \frac{1}{2} (\mathrm{i}+\omega^{(k)}_l)^{-1}y_k^2}\\
\end{split}
\end{equation}
As by definition this must be proportional to 
$\exp(-\omega_k^{(j)}y_k^2/2)$, it follows that
\begin{equation}\label{eq:B5}
\omega^{(j)}_k=\frac{\mathrm{i}\lambda_{\varepsilon}}{r_k} + \sum_{l\in \partial k \setminus j} \frac{\mathrm{i}\omega_l^{(k)}}{\mathrm{i}+\omega_l^{(k)}}
\end{equation}
which after replacing $r_k=(\tau_k c)^{-1}$ is  the desired cavity equation. The calculation for the marginal precisions $\{\omega_{j}\}$ (see (\ref{Precisions})) is exactly analogous. Note that the discussion so far allows any kind of graph structure, i.e.\ it is independent of a specific choice for the degree distribution $p_k$, and it also allows correlations between degree and energies.

When going from the above considerations for a single finite-sized graph to the thermodynamic limit, one assumes that the cavity precisions $\{\omega_k^{(j)}\}$ are random variables taken from some distribution $p(\omega)$. Equation (\ref{eq:B5}) then turns into a self-consistent equation for $p(\omega)$. For a general degree distribution $p_k$ and a joint distribution $\rho_{\tau,k}(\tau,k) = \rho_{\tau|k}(\tau|k)p_k$ this reads

\begin{equation}\label{eq:B6}
p(\omega)=\sum_k \frac{p_k k}{c} \int \mathrm{d}\tau\,\rho_{\tau|k}(\tau|k)\prod_{l=1}^{k-1}\mathrm{d}\omega_{l}\,p(\omega_{l})\,\delta(\omega - \Omega_{k-1})
\end{equation}
where $c$ is the average degree of the network, $p_k k/c$ is the probability that a randomly chosen edge connects the root-node to a neighbour with degree $k$, and
\begin{equation}\label{eq:B7}
\Omega_{k-1} = \Omega_{k-1} (\{\omega_l\},\tau)= \mathrm{i}\lambda_{\varepsilon} \tau c + \sum_{l=1}^{k-1}\frac{\mathrm{i}\omega_{l}}{\mathrm{i}+\omega_{l}}
\end{equation}
Clearly (\ref{eq:B6}) reduces to the result for random regular graphs (\ref{Zeta}) in the main text once we impose that $\rho_{\tau|k}(\tau|k)=\rho_{\tau}(\tau)$ and $p_k=\delta_{c,k}$. The population dynamics algorithm for the general case (\ref{eq:B6}) follows the same protocol as discussed in section \ref{section4}, with the only difference that, at each update, one has to pick $k$ randomly with weight $p_k k/c$, then draw $k-1$ elements from $\mathcal{P}$ and $\tau$ from $\rho_{\tau|k}(\tau|k)$.

We conclude this appendix with a final remark: while the change of variable $y_i=x_i r_i^{1/2}$ is not essential for single instance cavity evaluations, i.e.\ for fixed realisations of the disorder, this step becomes necessary in going to the thermodynamic limit. This is because otherwise correlations between cavity precisions on different branches of a cavity graph would be created by the coupling to the local disorder, and therefore the assumption of statistical independence between these cavity precisions would be violated.

%////////////////////////////////////////////////////////////////////////////////////////%

\section{High $T$ approximation}\label{appendixC}

\begin{figure}[htbp]
	\centering
	\includegraphics[height=3.5cm]{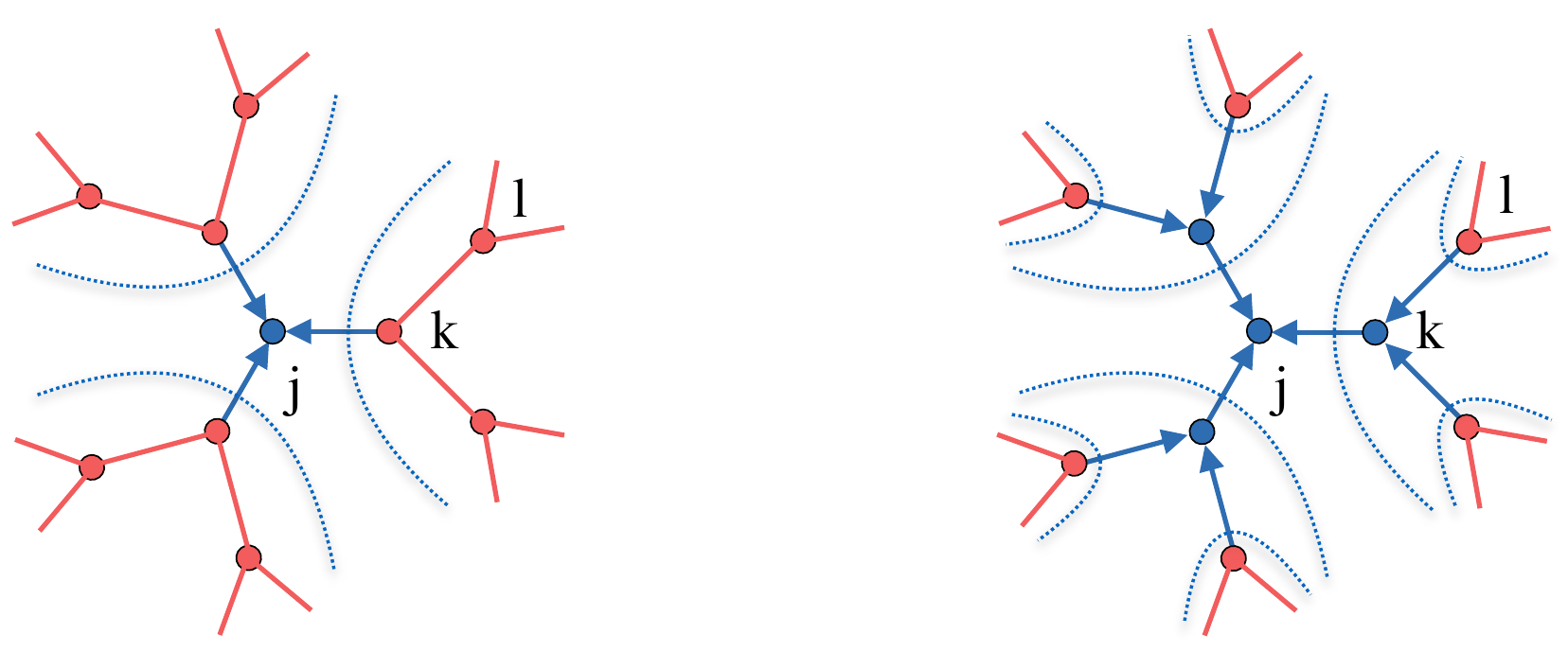}
	\caption{Left: schematic representation of the first order approximation: the sub-graph in red (only a small portion is shown here, namely the nearest and next-nearest neighbours of $j$), where disorder is absent, follows the infinite temperature solution. The ``messages'' from the nearest neighbour nodes $k$ feed into the central node through one cavity step at finite $T$ (blue arrows indicate an evaluation involving energy disorder). Right: at the second order we take two finite $T$ steps starting from the infinite temperature solution at the next-nearest neighbours $l$. }\label{fig:Approximation}
\end{figure}

In this appendix we discuss the construction of the high $T$ approximation for the DOS, $\rho^{A}(\lambda)$ (see (\ref{DOS_approximation})). As mentioned in section \ref{section4}, we take one cavity iteration at finite $T$, starting from the infinite temperature solution. Consequently the cavity precisions are evaluated without on-site disorder as the limit $T\to \infty$ gives $\tau_k=1$ for all $k$. Also, in the thermodynamic limit the random regular graph structure becomes effectively a regular tree, and the problem of finding the cavity precisions becomes free of disorder. Figure \ref{fig:Approximation}-left shows a schematic representation of this procedure. In this non-disordered framework the distribution $p(\omega)$ is then expected to be delta peaked on some value $\bar{\omega}$. We have
\begin{equation}\label{eq:C1}
\begin{split}
p(\omega) = \delta(\omega-\bar{\omega}) & =\int \mathrm{d}\tau \rho_{\tau}(\tau)\prod_{l=1}^{c-1} \mathrm{d}\omega_l \,p(\omega_l)\delta(\omega-(\mathrm{i}\lambda_{\varepsilon}c+\sum_{l=1}^{c-1}\frac{\mathrm{i}\omega_l}{\mathrm{i}+\omega_l}))\\
& = \int\prod_{l=1}^{c-1} \mathrm{d}\omega_l \,\delta(\omega_l-\bar{\omega})\delta(\omega-(\mathrm{i}\lambda_{\varepsilon}c+\sum_{l=1}^{c-1}\frac{\mathrm{i}\omega_l}{\mathrm{i}+\omega_l}))\\
& = \delta(\omega-(\mathrm{i}\lambda_{\varepsilon}c+(c-1)\frac{\mathrm{i}\bar{\omega}}{\mathrm{i}+\bar{\omega}}))
\end{split}
\end{equation}\label{eq:C2}
Thus $\bar{\omega}$ needs to satisfy
\begin{equation}
\bar{\omega}=\mathrm{i}\lambda_{\varepsilon}c+(c-1)\frac{\mathrm{i}\bar{\omega}}{\mathrm{i}+\bar{\omega}}
\end{equation}
and out of the two roots we need to pick the one with $\mathrm{Re}\,\bar{\omega}\geq 0$, to which we also simply refer as $\bar{\omega}$. We recall that the DOS in the thermodynamic limit is obtained by averaging over the cavity precisions and waiting time distributions according to
\begin{equation}\label{eq:C3}
\rho(\lambda) =\lim_{\varepsilon \to 0} \frac{1}{\pi} \text{Re}\Big\langle \frac{\tau c}{\Omega_c(\{\omega_l\},\tau)} \Big\rangle_{\{\omega_l\},\tau}
\end{equation}
The first order of our approximation scheme thus gives
\begin{equation}\label{eq:C4}
\rho^{A}(\lambda) =\lim_{\varepsilon \to 0} \frac{1}{\pi} \text{Re}\Big\langle \frac{\tau c}{\Omega_c(\{\bar{\omega}\},\tau)} \Big\rangle_{\tau}
\end{equation}
where
\begin{equation}\label{eq:C5}
\begin{split}
\Big\langle \frac{\tau c}{\Omega_c(\{\bar{\omega}\},\tau)} \Big\rangle_{\tau} & =\int_1^{\infty} \mathrm{d}\tau \,\rho_{\tau}(\tau) \frac{\tau}{\mathrm{i}\lambda_{\varepsilon}\tau +\frac{\mathrm{i}\bar{\omega}}{\mathrm{i+\bar{\omega}}}}\\
& = \frac{T}{\mathrm{i}\lambda_{\varepsilon}}\int_1^{\infty} \mathrm{d}\tau\, \frac{\tau^{-T}}{\tau+C(\lambda_{\varepsilon})}
\end{split}
\end{equation}
with $C(\lambda_{\varepsilon})=\bar{\omega}/(\lambda_{\varepsilon}(\mathrm{i+\bar{\omega}}))$. 
The integral over $\tau$ can be done directly giving a hypergeometric function of $T$ and $C$. Explicitly, one obtains
\begin{equation}\label{eq:C6}
\rho^A(\lambda)=\text{Re}\big[{}_2F_1(1,T;1+T\,|-C(\lambda))/\mathrm{i}\pi\lambda\big]
\end{equation}

The second order approximation consists of two cavity steps at finite $T$ -- and similarly the $n^{\text{th}}$ order approximation would have $n$ cavity steps -- starting from the infinite temperature solution (see figure \ref{fig:Approximation}-right). The cavity precisions are evaluated with their on-site disorder. As a result, the average giving the DOS contains the disorder of the neighbouring environment $\{\tau_k\}$, plus the local disorder of the central node $\tau$: 
\begin{equation}\label{eq:C7}
\rho^{\mathrm{2A}}(\lambda)=\lim_{\varepsilon\to 0}\frac{1}{\pi}\text{Re}\Big\langle \frac{\tau c}{\Omega_c(\{\Omega_{c-1}(\{\bar{\omega}\},\tau_k)\},\tau)} \Big\rangle_{\{\tau_k\},\tau}
\end{equation}
Similarly to (\ref{eq:C5}) we have
\begin{equation}\label{eq:C8}
\begin{split}
\Big\langle \frac{\tau c}{\Omega_c(\{\Omega_{c-1}(\{\bar{\omega}\},\tau_k)\},\tau)} \Big\rangle_{\{\tau_k\},\tau} & =\int \mathrm{d}\tau \,\rho_{\tau}(\tau) \prod_{k=1}^{c} \mathrm{d}\tau_k\,\rho_{\tau}(\tau_k) \frac{\tau c}{\mathrm{i}\lambda_{\varepsilon}\tau c+\sum_{k=1}^{c}\frac{\mathrm{i}\Omega_{c-1}(\{\bar{\omega}\},\tau_k)}{\mathrm{i+\Omega_{c-1}(\{\bar{\omega}\},\tau_k)}}}
\end{split}
\end{equation}
The results of numerical evaluation of $\rho^{2A}(\lambda)$ are discussed in section \ref{section4}.

Since the equation for $\bar{\omega}$ is $T$-independent, the localisation transitions detected by this approximation do not depend on temperature and they always lie at the ends of the RW-limit spectrum, specifically for the RRG case we have $\lambda^{\mathrm{L/R}}=-(2c-1)/c, -1/c$ (from (\ref{RW_DOS})). This is true for any finite iteration of the approximation scheme, as can be argued inductively: the $m$-th iteration cavity precisions will be imaginary -- indicating a localised region of the spectrum -- whenever they are evaluated using an imaginary $(m-1)$-th iteration solution, as long as the imaginary part in $\lambda_{\varepsilon}$ used for these evaluations is kept small enough.

%////////////////////////////////////////////////////////////////////////////////////////%

\section{The inverse participation ratio $I_2^{\star}$}\label{appendixD}

In this appendix we explain how to derive the equation (\ref{IPR2}), which constitutes an alternative to the formula that Boll\'e \emph{et al} proposed for estimating the average IPR \cite{Metz2010}. We start by expressing the diagonal resolvent entries in terms of the eigenvector components
\begin{equation}
G_{jj} (\lambda_{\varepsilon}) = \sum_{\alpha=1}^N \frac{v_{\alpha,j}^2}{\lambda-\mathrm{i}\varepsilon + \lambda_{\alpha}}
\end{equation}
whose imaginary part reads
\begin{equation}
\text{Im}\,G_{jj} (\lambda_{\varepsilon}) = \sum_{\alpha=1}^N \frac{\varepsilon}{(\lambda-\lambda_{\alpha})^2 + \varepsilon^2}\, v_{\alpha,j}^2
\label{Im_G}
\end{equation}
In order to simplify the notation we will omit the lambda argument in $G_{jj}(\lambda_{\varepsilon})$ and $\rho(\lambda)$ where necessary. We also take $N$ as large but finite and assume that $\rho(\lambda)$ is finite, too. This can be ensured by choosing $\varepsilon$ small but such that $N\varepsilon \gg 1$: for any given $\lambda$ many eigenvalues then contribute to $\rho(\lambda)$, which 
evaluates the DOS using Lorentzians $\delta_{\varepsilon}(\lambda - \lambda_{\alpha})$ of width $\varepsilon$. Since the statistics of the $\text{Im}\, G_{jj}$ are crucial in determining the value of important quantities like the DOS, we consider the associated cumulant generating function $F({\bf q}) = \ln \langle \text{exp}(\sum_j q_j\,\text{Im}\,G_{jj})\rangle$, %_{G_{jj}}
which can be expressed as
\begin{equation}\label{CMF}
F({\bf q}) \simeq \Biggl\langle \ln \prod_{\alpha,j} \big\langle e^{q_j\varepsilon v^2 / [(\lambda-\lambda_{\alpha})^2 +\varepsilon^2]} \big\rangle_v \Biggr\rangle_{\{\lambda_\alpha\}}
\end{equation}
Here we have made two approximations. The first is to treat the $v_{\alpha,j}$ for different $\alpha$ and $j$ as independent, thus ignoring normalisation and orthogonality constraints on the eigenvectors. This is plausible as the number of constraints is much smaller than the number of variables $v_{\alpha,j}$, producing only weak correlations. To see this, note that there are $O(N\rho \varepsilon)$ eigenvectors contributing significantly to (\ref{Im_G}). These have $O(N\times N\rho\varepsilon)$ components, while the number of orthonormality constraints between them is $O(N^2\rho^2\varepsilon^2)$ and so smaller by a factor $\varepsilon$.
The second approximation in (\ref{CMF}) is that we are neglecting correlations between eigenvalues and eigenvectors. This again seems plausible given that the eigenvalues $\lambda_\alpha$ that contribute lie within a small range of $O(\varepsilon)$ around $\lambda$ where the statistics of the $v_{\alpha,j}$ should change little.

To evaluate $F({\bf q})$ it now remains to average over the $\lambda_\alpha$. As the number of contributing eigenvalues is $O(N\rho\varepsilon)$ and hence large, small fluctuations of the eigenvalues around their mean positions should be immaterial. We therefore approximate the $\lambda_\alpha$ as lying on a linear grid with the relevant spacing $(N\rho)^{-1}$, and in the same spirit replace the sum over $\alpha$ by an integral, shifting its origin so that $\alpha=0$ designates the eigenvalue closest to $\lambda$. Then 
(\ref{CMF}) becomes
\begin{equation}\label{CMF2}
F({\bf q}) \simeq \sum_j 
\int d\alpha\, \ln\big\langle e^{q_j\varepsilon v^2 / [(\alpha/(N\rho))^2 +\varepsilon^2]} \big\rangle_v = 
\sum_j 
\int d\alpha\, \ln\big\langle e^{\beta_\alpha(q_j)v^2} \big\rangle_v
\end{equation}
where we have defined
\begin{equation}
\beta_{\alpha}(q) = \frac{N^2\rho^2\varepsilon q}{\alpha^2 + N^2\rho^2\varepsilon^2}
\end{equation}
Expanding in the $q_j$ now gives a conventional cumulant expansion
\begin{equation}
F({\bf q}) \simeq 
%\sum_{\alpha}\ln \langle e^{\beta_{\alpha}(q) v^2} \rangle=
\sum_{j,n} \int d\alpha\,\frac{\beta_{\alpha}^n(q_j)}{n!}\mathcal{K}_{v^2}^n
\end{equation}
where we denote by $\mathcal{K}^{n}_x$ the $n$-th cumulant of $x$. The remaining integral is
\begin{equation}
\int d\alpha\,\beta_{\alpha}^n(q)=
q^n N \rho
\varepsilon^{-(n-1)}c_n
\end{equation}
where $c_n = \int \mathrm{d}x (1/(1+x^2))^n$, so that
\begin{equation}
F({\bf q}) \simeq 
\sum_{j,n} \frac{q_j^n}{n!}
N \rho
\varepsilon^{-(n-1)}c_n
\mathcal{K}_{v^2}^n
\end{equation}
Picking out the term of order $q_j^n$ finally leads to the following correspondence between the $n$-th cumulants of $\text{Im}\,G_{jj}$ and $v^2$
\begin{equation}\label{cumulants_n}
\mathcal{K}_{\text{Im}\,G_{jj}}^n \simeq N\rho(\lambda)\varepsilon^{-(n-1)}c_n \mathcal{K}_{v^2}^n
\end{equation}

We can now use the above general result to relate the second cumulant of $v^2$ to the IPR: from the definition (\ref{IPR}) we have $I_2(\lambda)\simeq\langle v^4\rangle/(N\langle v^2\rangle^2)$, while generally $\mathcal{K}_{v^2}^{2}=\langle v^4\rangle - \langle v^2 \rangle^2$. Imposing the eigenvector normalization condition $\langle v^2 \rangle = 1/N$ we obtain
\begin{equation}
\mathcal{K}_{v^2}^{2} \simeq \frac{1}{N^2} (NI_2(\lambda) - 1)
\end{equation}
When $I_2(\lambda) = \mathcal{O}(1)$, i.e.\ in a localised part of the spectrum, the first term dominates for large $N$ and we obtain $\mathcal{K}_{v^2}^{2}\simeq I_2(\lambda)/N$. Equation (\ref{cumulants_n}) evaluated at second order then gives a formula for the IPR that is $N$-independent:
\begin{equation}\label{IPR_2_G}
I_2(\lambda) = \lim_{\varepsilon \to 0} \lim_{N\to\infty}\frac{2\varepsilon}{\pi\rho(\lambda)}\text{Var}\big[\text{Im}\,G_{jj}\big]
\end{equation}
where we have noted explicitly the order of limits involved. In the large $N$-limit taken, 
one expects $I_2$ to be self-averaging and thus approach some value $\bar{I}^{\star}_2$. Substituting (\ref{Resolvent_entries_precision}) into (\ref{IPR_2_G}), in the infinite-$N$ limit we finally obtain the IPR estimate (\ref{IPR2}) in the main text.

\begin{figure}[h]
	\centering
	\includegraphics{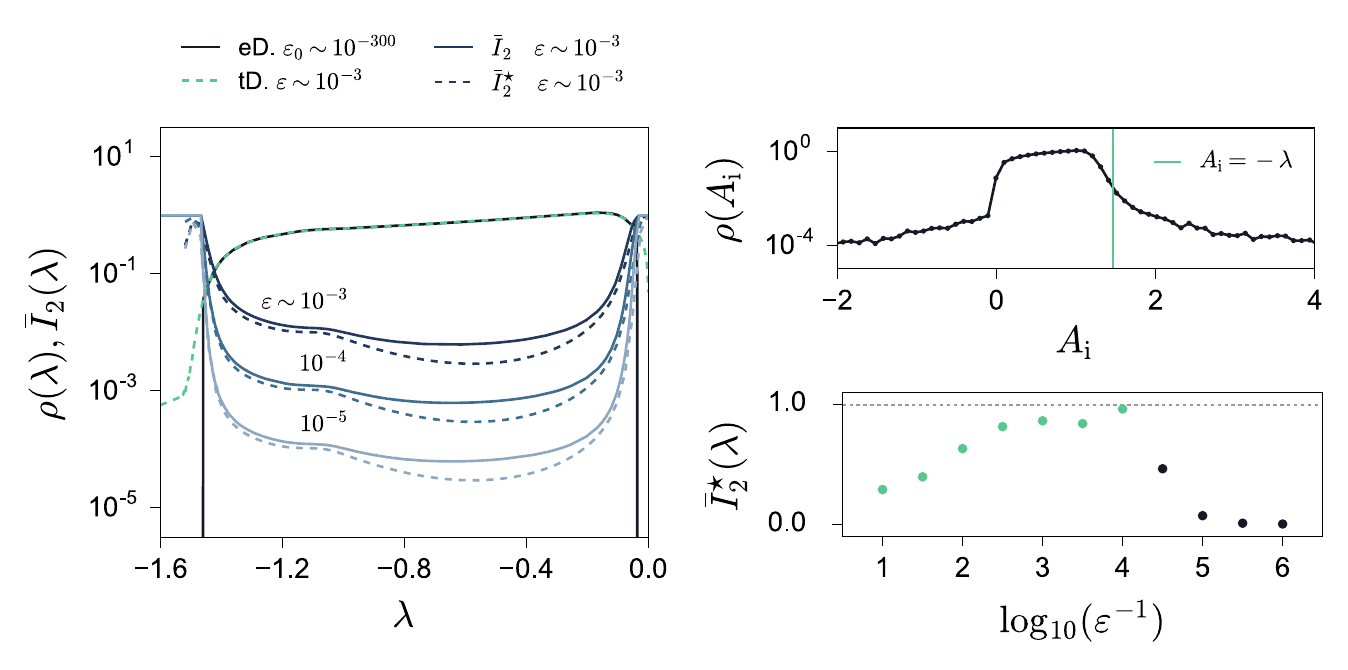}
	\caption{Left: average IPR evaluated via (\ref{IPR})  and (\ref{IPR2})  using different values of $\varepsilon$ (dark blue to light blue), extended DOS (black) and total DOS (green dashed line); $\bar{I}_2^{\star}(\lambda)$ from (\ref{IPR2}) is averaged within $\lambda$-bins for clearer visualisation. In the extended region of the spectrum the IPR scales with $\varepsilon$ as expected. Top right: histogram of the $A_{\mathrm{i}}$ values collected at $\lambda \simeq -1.465$ (in the localised region on the left side of the spectrum), note that $\rho(A_{\mathrm{i}})$ is smooth around $A_{\mathrm{i}}=-\lambda$. Bottom right: $\bar{I}_2^{\star}(\lambda)$ against decreasing values of $\varepsilon$. The green points converge to the limiting value of unity, the black points drop to zero because the $\varepsilon$ values used over there are too small to ensure proper averaging. Evaluations were performed using a population of size $N_{\mathrm{p}}=2500$, with temperature and connectivity of $T=1.5$ and $c=5$, respectively.
	}\label{fig:IPRs}
\end{figure}

Figure \ref{fig:IPRs}-left shows the IPR estimated via (\ref{IPR}) and (\ref{IPR2}) using different values of $\varepsilon$, at temperature $T=1.5$ and for average connectivity $c=5$; the total and extended DOS are also included. We observe that the IPR scales linearly with $\varepsilon$ within the bulk of the spectrum, which is as expected for (\ref{IPR}). Our alternative estimate (\ref{IPR2}) is directly applicable only within the localised part of the spectrum, but also turns out to be $O(\varepsilon)$ for extended states. 
In finite systems, the IPR for extended states is $O(1/N)$. Intuitlvely, one can therefore say that
in the population dynamics algorithm, which assumes $N\to\infty$, the ``regulariser'' $\varepsilon$ effectively plays the role of the inverse system size, $1/N$.

In the localised part of the spectrum, figure \ref{fig:IPRs}-left shows that both IPR estimates are of order unity, though $\bar{I}_2(\lambda)=1$ throughout for $\varepsilon\to 0$ while $\bar{I}_2^{\star}(\lambda)$ remains below unity as one would expect physically (an average IPR of one would require all eigenvectors to be localised onto a single node, which is not even true in the mean-field limit). We note, however, that $\bar{I}_2^{\star}(\lambda)$ can be written as
\begin{equation}\label{IPR2_bis}
\bar{I}_2^{\star}(\lambda) = \lim_{\varepsilon\to 0} 2\varepsilon\frac{\langle x^2\rangle - \langle x\rangle^2}{\langle x\rangle}
\end{equation}
where, using the notation of (\ref{IPR_A}),
\begin{equation}
x = \frac{\varepsilon+A_{\mathrm{r}}}{(\varepsilon+A_{\mathrm{r}})^2+(\lambda+A_{\mathrm{i}})^2}
\end{equation}
The second (mean squared) term in (\ref{IPR2_bis}) is irrelevant in the limit $\varepsilon\to 0$, because $\langle x\rangle=\pi\rho(\lambda)$ is of order unity.
In the localised part of the spectrum $A_{\mathrm{r}}=0$ so $x$ simplifies to $x=\varepsilon/(\varepsilon^2+(\lambda+A_{\mathrm{i}})^2)$. Here, the remaining term
\begin{equation}\label{IPR2_bis2}
\bar{I}_2^{\star}(\lambda) = \lim_{\varepsilon\to 0} 2\varepsilon\frac{\langle x^2\rangle}{\langle x\rangle}
\end{equation}
makes clear that the second moment of $x$ must be $O(1/\varepsilon)$ and hence significantly larger than the squared mean. In fact, if the distribution of $A_{\mathrm{i}}$ approaches a smooth limit $\rho(A_{\mathrm{i}})$ for $\varepsilon\to 0$, then (for small $\varepsilon$)
\begin{equation}\label{x_moments}
\langle x\rangle = \pi \rho(-\lambda), \qquad
\langle x^2\rangle = \frac{\pi}{2\varepsilon} \rho(-\lambda)
\end{equation}
because both are given by integrals that are sharply peaked at $A_{\mathrm{i}}=-\lambda$. Finally (\ref{IPR2_bis2}) and (\ref{x_moments}) give
\begin{equation}\label{IPR2_bis3}
\bar{I}_2^{\star}(\lambda) = 1
\end{equation}
This argument is confirmed by the data shown in figure \ref{fig:IPRs}-right: on the top we can clearly see that the histogram of $A_{\mathrm{i}}$ is smooth everywhere, and in particular around the value of $-\lambda$ that dominates the computation. The plot at the bottom shows the convergence (green points) of $\bar{I}_2^{\star}$ when epsilon decreases; here we have used the same data as in the plot for $\rho(A_{\mathrm{i}})$. The reason why the last few points (in black) drop to zero is that the amount of data collected was enough to give a smooth histogram on a scale as small as $10^{-4}$, but not less. 

In conclusion, our estimate for the average IPR is expected to give a value of unity in the localised region of the spectrum, in the limit $\varepsilon\to 0$. Surprisingly this is the same result given by the Boll\'e {\em et al} formula, even though the latter is based on the opposite assumption of the estimate of the IPR being dominated by a single eigenvector.
 
%////////////////////////////////////////////////////////////////////////////////////////%

\section{Numerical results for the IPR}\label{appendixE}

\begin{figure}[h]
	\centering
	\includegraphics{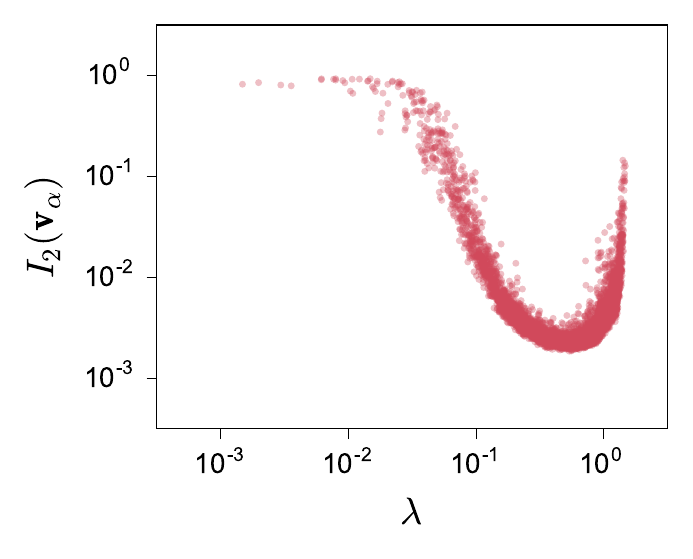}\includegraphics{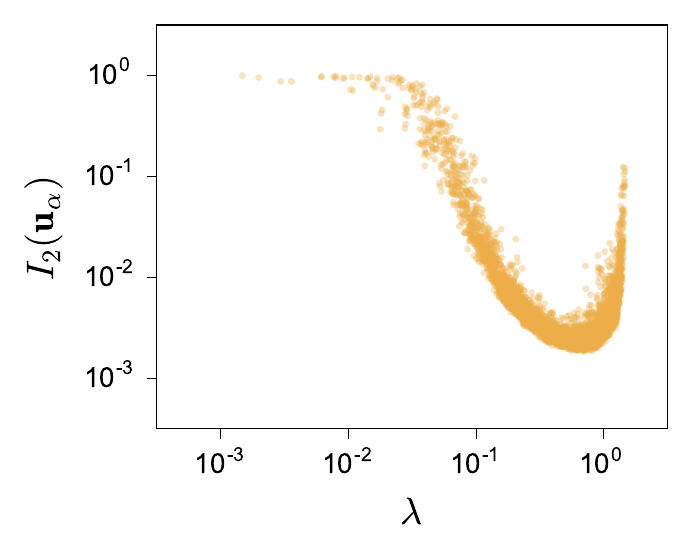}\\\includegraphics{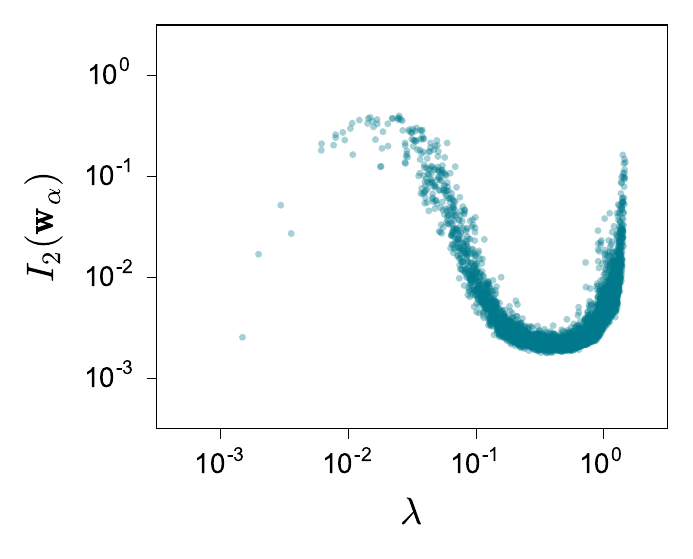}\includegraphics{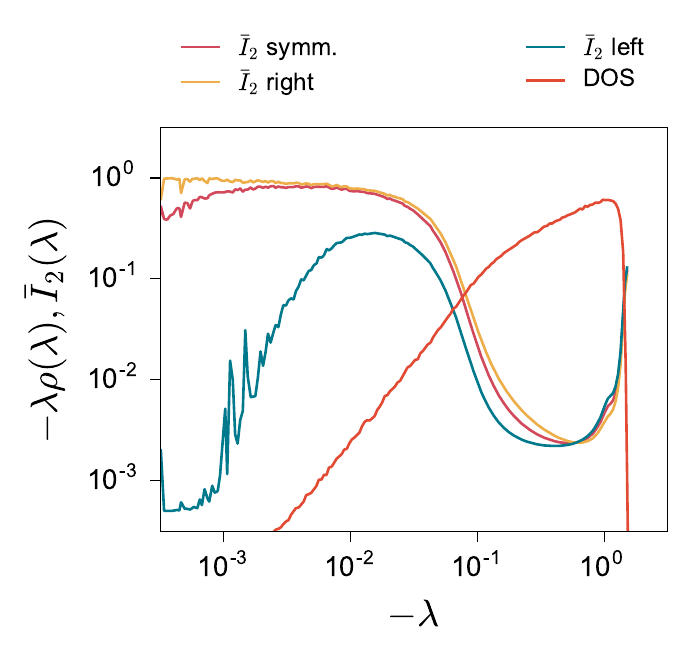}
	\caption{Scatterplots of IPR values of the symmetrised $\mathbf{v}_{\alpha}$, right $\mathbf{u}_{\alpha}$ and left $\mathbf{w}_{\alpha}$ eigenvectors against $r=-\lambda$ for the case of random regular graphs with system parameters $c=5$, and $T=1.5$. The data have been collected across $M=2000$ samples of size $N=2000$. The bottom-right plot shows the bin-wise average of the IPR values in the scatterplots, together with the DOS.}\label{fig:numerical_IPR}
\end{figure}

\begin{figure}[h]
	\centering
	\includegraphics{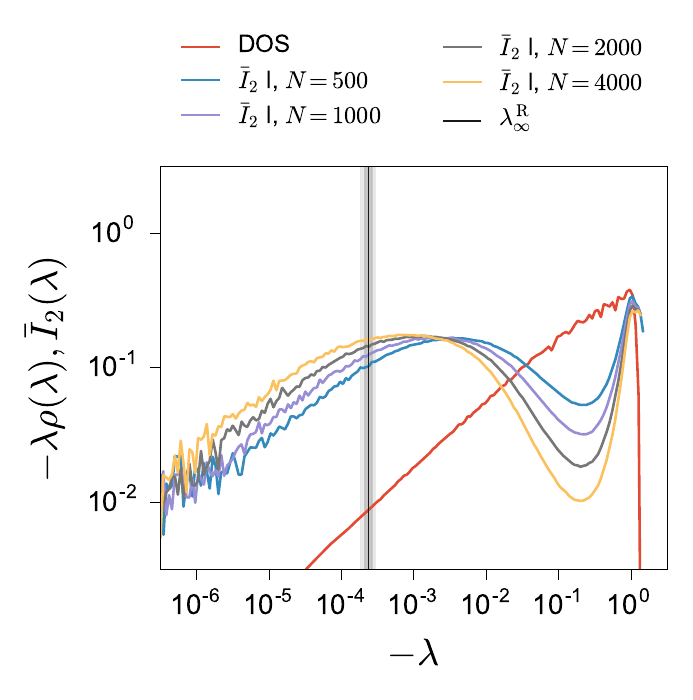}\includegraphics{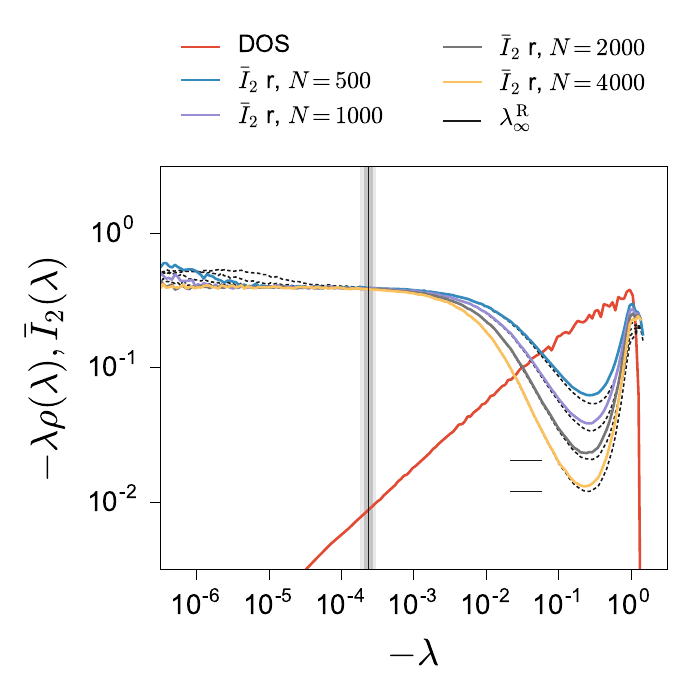}
	\caption{The plots show the DOS and the bin-wise average IPR of the left $\mathbf{w}_{\alpha}$ (left plot), right $\mathbf{u}_{\alpha}$ (right plot) and symmetrised $\mathbf{v}_{\alpha}$ (black dashed line in the right plot) eigenvectors for the case of random regular graphs with system parameters $c=5$, and $T=0.5$, and system size $N=500, 1000, 2000, 4000$. The vertical black line represents the extrapolated mobility edge predicted from population dynamics (i.e.\ in the infinite system size limit), with shaded areas covering the $68\%$ and $95\%$ confidence interval. The horizontal black lines in the right plot show the decrease of the IPR by a factor of two when $N$ is increased from 2000 to 4000.}\label{fig:numerical_IPR_T05}
\end{figure}

In section \ref{section3} we discussed briefly the effect of the symmetrisation (\ref{Master_operator}) on the localisation properties of the eigenvectors, focussing particularly on the ground state. Recall that the right, left and symmetrised eigenvectors, respectively $\mathbf{u}_{\alpha}$, $\mathbf{w}_{\alpha}$ and $\mathbf{v}_{\alpha}$, are related via $\mathbf{P}_{\mathrm{eq}}$ as $\mathbf{v}_{\alpha} = \mathbf{P}_{\mathrm{eq}}^{-1/2} \mathbf{u}_{\alpha} = \mathbf{P}_{\mathrm{eq}}^{1/2} \mathbf{w}_{\alpha}$, where $(\mathbf{P}_{\mathrm{eq}})_{ii}=p_i^{\mathrm{eq}}\propto \tau_i = \mathrm{exp}(\beta E_i)$ and $(\mathbf{P}_{\mathrm{eq}})_{ij}=0$. In the infinite temperature limit (RW) the symmetrisation is immaterial as $\mathbf{P}_{\mathrm{eq}}$ reduces to the identity matrix, which implies $\mathbf{v}_{\alpha} = \mathbf{u}_{\alpha} = \mathbf{w}_{\alpha}$. In appendix \ref{appendixA} we have discussed the localisation in the mean field limit where the system has no spatial structure and the IPR of symmetrised or non-symmetrised eigenvectors is dominated by the pole in $-\lambda$ (see equations (\ref{MF_eigenvector_element}, \ref{MF_eigenvector_element_symmetric})); the factor $\tau_i^{-1/2}$ in the numerator of the symmetrised case does not affect the value of the IPR qualitatively. Likewise, we generally expect that multiplying element-wise the eigenvectors by a smooth function of the energy will not change the qualitative behaviour of $I_2(\lambda)$. This idea is confirmed by the numerical results presented in this appendix. Figure \ref{fig:numerical_IPR} shows the IPR of left, right and symmetrised eigenvectors across the entire $\lambda$-range (except for the ground state $\lambda = 0$) for the case of random regular graphs with mean connectivity $c=5$ and temperature $T=1.5$. We observe that the different choices of eigenvectors have qualitatively the same localisation behaviour, except in the range of small $r=-\lambda$, where there is a natural crossover to the ground state value. We also observe that the IPR values in the scatterplots are mostly concentrated on their bin-wise average (bottom-right). The latter also indicates that symmetric and right eigenvectors have almost overlapping values of $I_2$. Figure \ref{fig:numerical_IPR_T05} shows the average IPR of left, right and symmetric eigenvectors across the $\lambda$-range, for different values of the system size $N$, together with the DOS and the extrapolated mobility edge from population dynamics (see section \ref{section4}).  The $N$-dependence is as expected: the average IPR scales as $1/N$ in delocalised regions and it is of order $1$ for localised regions. The scaling with $1/N$ is illustrated by the horizontal lines in the right plot. These are separated by a factor of 2 on the $y$-axis corresponding to the change from $N=2000$ to $N=4000$. The separation on the $r$-axis between the two regimes (localised and delocalised) is consistent with the extrapolated mobility edge (see particularly the right plot), though the decrease of the IPR towards values of order $1/N$ is slow in the delocalised regime near the mobility edge.

%////////////////////////////////////////////////////////////////////////////////////////%

\end{document}